\documentclass[a4paper, 12pt]{article}

\usepackage{graphicx}
\usepackage{txfonts}

\title{Heating of braided coronal loops}

\author{A.L. Wilmot-Smith \footnote{Division of Mathematics, University of Dundee, 
Dundee, DD1 4HN, United Kingdom.}, 
D.I. Pontin $^{*}$, 
A.R. Yeates $^{*}$ \footnote{Present address: Department of Mathematical 
Sciences, Durham University, Durham, DH1 3LE.} \ , 
G. Hornig $^{*}$.}

\date{October 2011}

\begin{document}

\maketitle

\abstract{
We investigate the relaxation of braided magnetic loops in order to find out how the type of braiding
via footpoint motions affects resultant heating of the loop.
Two  magnetic loops, braided in different ways, are used as initial conditions in 
resistive MHD simulations and their subsequent evolution is studied. 
The fields both undergo a resistive relaxation in which current sheets form and fragment
and the system evolves towards a state of lower energy.   
In one case this relaxation is very efficient with current sheets filling the volume and 
homogeneous heating of the loop occurring.  In the other case fewer current sheets develop, less 
magnetic energy is released in the process and a patchy heating of the loop results. 
The two cases, although very similar in their setup, can be distinguished by the mixing properties 
of the photospheric driver.  The mixing can be measured by the topological entropy of 
the plasma flow, an observable quantity.}

\begin{center}  
Sun: corona;  Sun: magnetic topology.
 \end{center}

\section{Introduction}
\label{sec:intro}

Coronal loops are enormously diverse in their nature, acting as building blocks of the corona, 
from bright points to active regions and flaring loops.  As such, loops cover a huge range of 
lengths ($1-1000$ Mm) and it seems likely that several coronal heating mechanisms are responsible 
for heating loops to the observed range of temperatures  ($0.1 - > \!\!10$ MK). 
Explaining the observations remains a challenge and a number of questions are currently 
under debate.  For example, can loops be broadly 
classified as isothermal or multi-thermal  (e.g.~Schmelz et al.~2009; Aschwanden \& Boerner 2011)?
Is heating impulsive or steady (e.g.~Patsourakos \& Klimchuk 2008; 
Tripathu {\it et al.}~2010;  Warren  {\it et al.}~2010)?  A recent review of observations and modelling of 
coronal loops can be found in Reale (2010).  Furthermore, a more general coronal heating review is 
given by Klimchuk (2006).

One very promising loop-heating method, following the early ideas of Gold (1964) and Parker (1979, 1994),  
is magnetic braiding.   Here photospheric motions acting on the loop footpoints act to twist and tangle the 
overlying field, increasing its magnetic energy.  Eventually current layers (singular or non-singular) may form 
in the field  (e.g.~Longcope \& Sudan 1994; Galsgaard \& Nordlund 1996; Longbottom {\it et al.}~1998; 
Ng \& Bhattacharjee, 1988; Craig \& Sneyd 2005; Wilmot-Smith {\it et al.}~2009).  Magnetic reconnection
will then enable a restructuring of the field as it relaxes to a lower energy state with plasma heating a natural 
consequence of the energy release.  Magnetic braiding is also a possible explanation for the observation that coronal loops have approximately 
constant width (Klimchuk 2000; L{\'o}pez Fuentez {\it et al.}~2008):  a braiding of field lines within a 
loop prevents the expansion seen in simple potential or linear-force--free models (L{\'o}pez Fuentez {\it et al.}~2006).

Although it is well-established that there is, in principle, sufficient energy in these photospheric motions for 
this mechanism to be plausible  (e.g.~Klimchuck 2006), 
whether the process is responsible for the coronal heating depends on many, 
often unkown, properties of the driver and the relaxation mechanism in the corona. One important 
question is how efficient the surface motions are at building up free energy in the magnetic field. 
Another unresolved issue is whether the relaxation mechanism in the corona can release this energy again. 
A theory often invoked to describe the energy release process is that of Taylor relaxation.
This theory was initially developed for a laboratory plasma device
(Taylor 1974, 1986) but has also been applied to the solar case (Heyvaerts \& Priest 1984,  Dixon {\it et al.}~1989, 
Nandy {\it et al.}~2003, Kusano 2005, Hood {\it et al.}~2009).
Under  this hypothesis,  the field relaxes to a particular linear force-free field (with the same global helicity, 
toroidal flux and boundary conditions as the initial field) so that not all the magnetic energy in 
excess of that of the potential field can be released.

The aim of this paper is to show that the amount of energy that can be released in any non-ideal
relaxation depends greatly on the topological properties of photospheric flow, which in turn determine  
the way  in which the magnetic field lines making up the flux tube are braided (mixed and tangled together). 
To show this we consider the resistive evolution of two contrasting magnetic fields, both generated through
 rotational motions on the boundary. The first field has a sequence of rotational footpoint motions of alternating sense.  
The comparison case  is also generated by rotational footpoint motions but these are all in the same sense.

 Our approach differs from  studies  (e.g.  Gudiksen \& Nordlund 2002; Peter {\it et al.}~2004;
 Bingert \& Peter 2011)  
 in which quasi-stationary processes of continuous driving and relaxation are modelled in that 
 we first assume an ideal process where the photospheric driver braids a coronal loop to a 
 certain level before  a resistive evolution is allowed.   This has the advantage that the more complex structures 
 we aim to study can be  built up without being quickly dissipated by high numerical resistivities.
  By contrast the present study involves a less sophisticated treatment of certain physical effects such as 
 heat conduction and radiative losses, with  results  not yet suitable for forward modelling.
  As such we view the two approaches as being complementary.

\begin{figure*}[tbp] 
\centering
\includegraphics[width=0.33\textwidth]{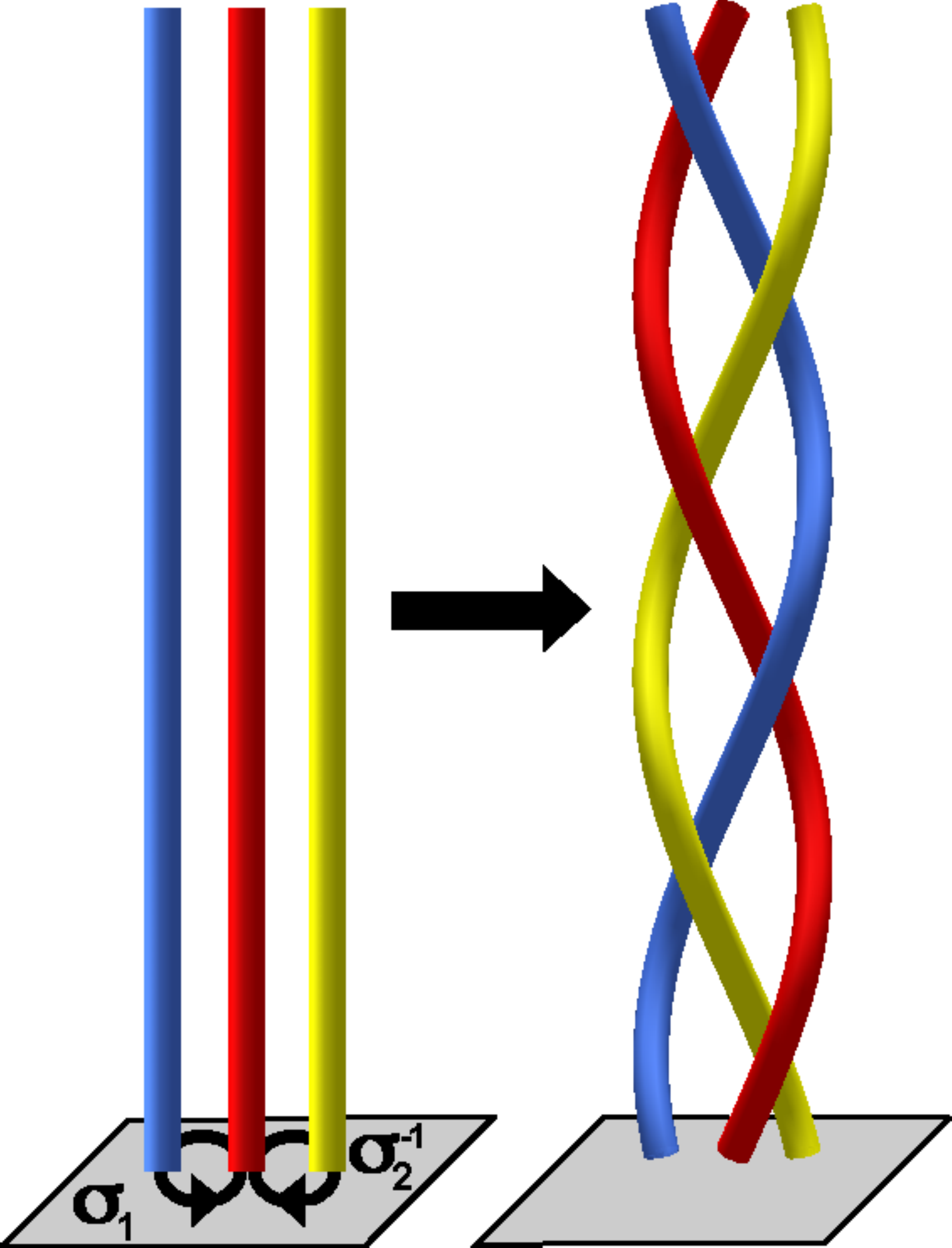}  \ \ \ \ \ \ \ \ \ \ \ \ 
\includegraphics[width=0.33\textwidth]{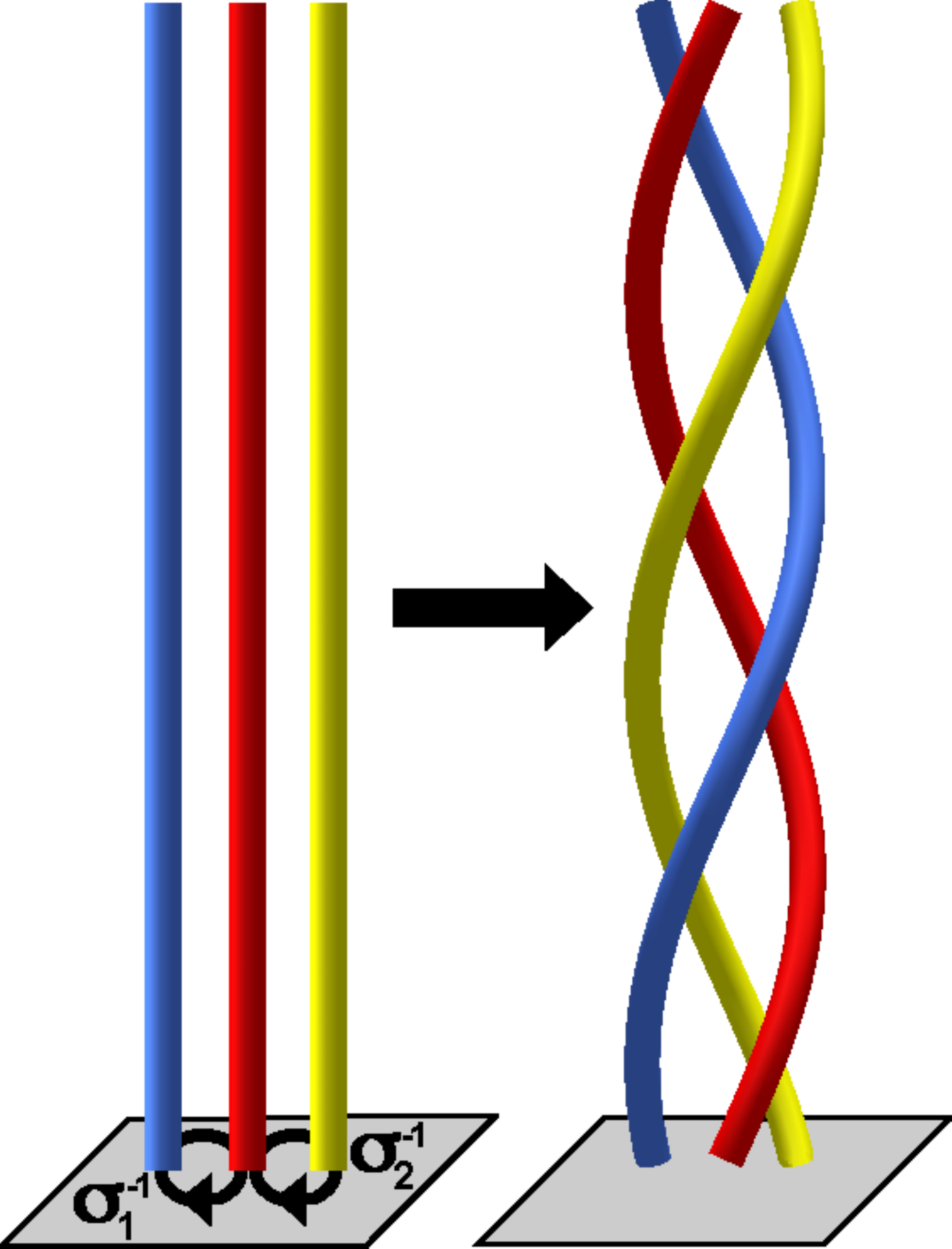} 
\caption{Cartoons showing the way in which the fields $E^{3}$ (left) and $S^{3}$ (right) can be 
built up by rotational footpoint motions acting on initially straight strands.}
\label{fig:braidcartoons}
\end{figure*}

An outline of the work is as follows:
In Section~\ref{sec:compare} we introduce the two magnetic fields ($E^{3}$ and $S^{3}$) 
whose MHD evolution will be studied throughout.  Tools for measuring the level of braiding are discussed.
In Section~\ref{sec:methods} we describe the numerical methods used for the simulations.
Results are presented in Section~\ref{sec:results} and are broken down into two parts, firstly
an examination of the basic properties of the resistive relaxation  and
secondly energetic considerations including estimates for coronal heating.
We conclude in Section~\ref{ref:conc}.

\section{Model Magnetic Loops}
\label{sec:compare}

 Throughout  this work, we employ an idealised representation of the coronal geometry, with 
nearly straight coronal fields running between two parallel places, which represent the photosphere.
 The first of our two magnetic fields, $E^{3}$, is based on the pigtail braid.  
It may be built up from a uniform vertical field  by rotational stirring motions on the boundary (the photosphere).   
The manner in which these motions act on uniform strands to create the braid
$E^{3}$ is illustrated in Figure~\ref{fig:braidcartoons} (left).
Two regions of rotational footpoint motion are present.  As viewed from the direction shown in the figure, the right-hand motion acts in a clockwise direction and the left-hand motion in an anti-clockwise direction.  Each motion rotates the strands about each other by a relative angle of $\pi$ radians.
  {The rotations are applied in the sequence 
$\sigma_{1}, \sigma_{2}^{-1},\sigma_1, \sigma_{2}^{-1},\sigma_1, \sigma_{2}^{-1}$,
which is also the braid word (Birman 1975)
 representing this braid (the power negative one indicates the change of orientation of the rotation). 
Since the  number of left and right hand rotations is the same the total magnetic helicity of the configuration 
is zero.  While the cartoon image shows just three field lines of the braid, the motions on the boundary will 
affect all field lines within a certain range and generate a  continuum of braid patterns. Those field lines lying 
outside the domain of the rotational motions will remain straight and undisturbed.}

Our second magnetic field, labelled $S^{3}$,  {is built up by a very similar sequence of motions 
on the boundary but with rotations all acting in the same, clockwise direction 
($ \sigma_{1}^{-1}, \sigma_{2}^{-1}, \sigma_{1}^{-1}, \sigma_{2}^{-1}, \sigma_{1}^{-1}, \sigma_{2}^{-1}$)}.  
The braid representation of this field 
is shown in Figure~\ref{fig:braidcartoons} (right) where we see that the corresponding three strands have each 
undergone exactly $2\pi$ rotation.  Again a continuum of  braiding patterns will be found in 
all field lines making up the loop but the total helicity of the field no longer vanishes.

We now construct an explicit magnetic field representation of these idealised pictures.
Assuming a Cartesian geometry, we take a uniform background field ($1 \mathbf{e}_{z}$) and 
superimpose six  flux rings, evenly spaced in $z$ and located alternately at $(x,y) = (1,0)$ or
 $(x,y) = (-1,0)$.  Each ring has components only in the 
$\mathbf{e}_{x}$ and $\mathbf{e}_{y}$ directions and is localised in all three dimensions.
Together this creates
six localised regions of twist in an otherwise uniform field.  The closed form expression which generates such 
magnetic fields is given in Wilmot-Smith {\it et al.} (2009).    For  $E^{3}$ we take exactly the same 
parameter set as given in that paper while for $S^{3}$ we simply change the sign of the twist parameter ($k$) 
to be the same ($k=1$) for each twist region (rather than alternating, as for $E^{3}$).

The aspect ratio of both loops is high ($1:8$).
 {This is broadly consistent with observations of coronal loops whose elementary strands
(width $ \lesssim 2$ Mm, Aschwanden 2005) are much longer than they are wide.}
 Both initial magnetic fields contain the same amount of  magnetic energy
($\int_{V} B^{2}/2\mu_0 \ dV$).  The braiding has been applied in a conservative manner with the magnetic energy 
being only a small amount ($3.08\%$) above that of the uniform background field.
In other respects the two braids are very different, as detailed in the following paragraphs.

\begin{figure}[tbp] 
\centering
\includegraphics[height=0.4\textwidth]{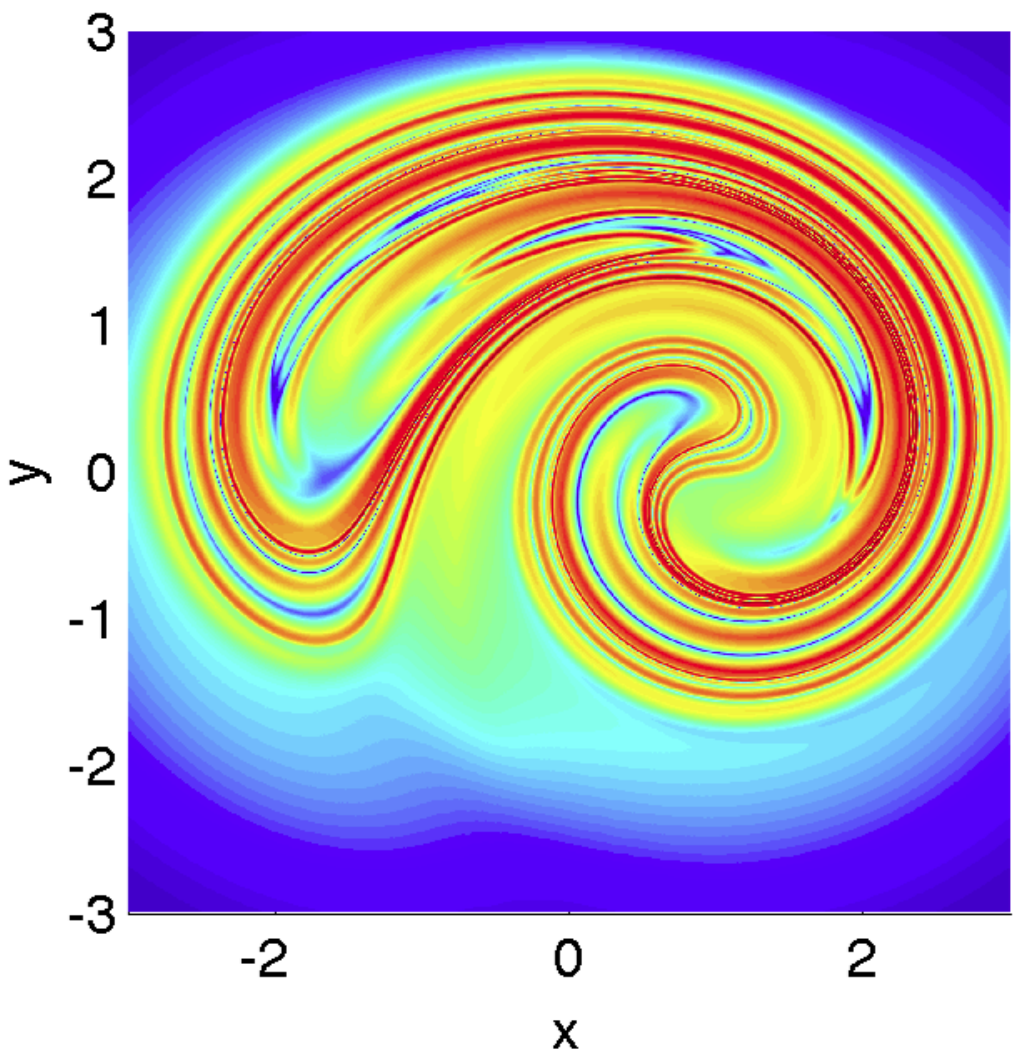} 
\includegraphics[height=0.4\textwidth]{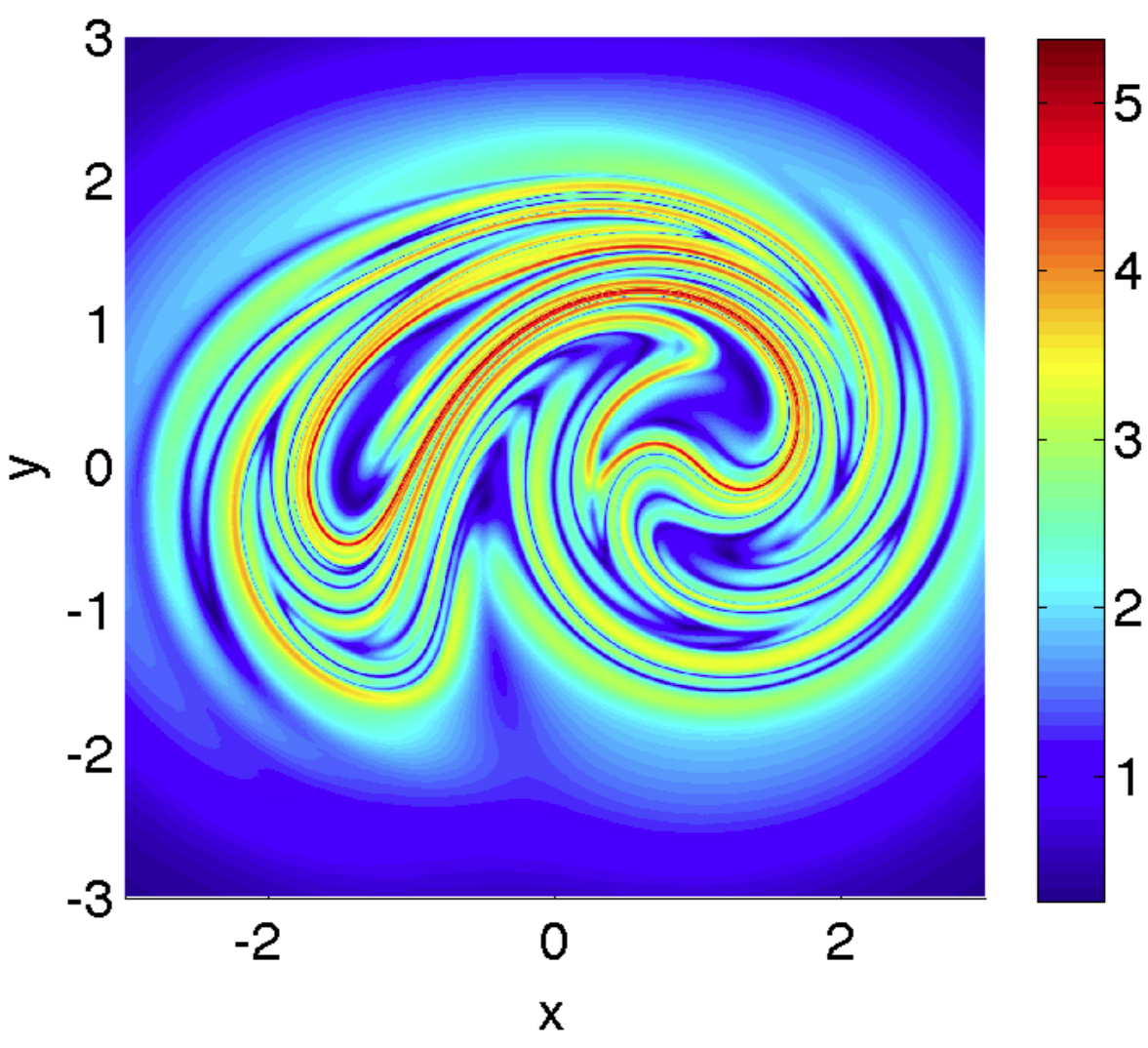} 
\caption{Distribution of $\log_{10}(Q)$ on the lower boundary for the initial state of $E^{3}$ (left) 
and $S^{3}$ (right).}
\label{fig:Qinitial}
\end{figure}

To understand these differences consider
the nature of the field line connectivities.  An established method for doing so 
is to examine the squashing factor $Q$ (Titov {\it et al.}~2002).
This is shown on the lower boundary of both fields in Figure~\ref{fig:Qinitial}.
While the maximum value of $Q$ in both fields is comparable 
(specifically, $Q(E^{3})_{max} = 2.3 \times 10^{5}$ and $Q(S^{3})_{max} = 2.4 \times 10^{5}$),
it is clear that there are many more layer-like regions of high $Q$ for $E^{3}$ than for $S^{3}$.
These regions arise from the property that field lines making up $E^{3}$ have a more complex 
connectivity than those of $S^{3}$,
there being many more regions in which neighbouring field lines diverge as they are traced up through 
the corresponding braids.  Overall this gives a `mixing' of field lines with respect to their connectivities on the
lower boundary. 
Simplistically, the efficiency of the field line mixing could be quantified by, for example, integrating $Q$ 
over the surface.  Calculating $\mathcal{Q} = \int_{A} \log_{10}(Q) dA$ 
(where $A$ is the surface $[-3,3] \times [-3,3]$ shown in Figure~\ref{fig:Qinitial})
for both fields we find $\mathcal{Q}_{E^{3}}=89.5$ while $\mathcal{Q}_{S^{3}}=68.0$.  These
values confirm the qualitative picture given in Figure~\ref{fig:Qinitial} that the field line mixing is
better for $E^{3}$.

\begin{figure}[tbp] 
\centering
\includegraphics[width=0.4\textwidth]{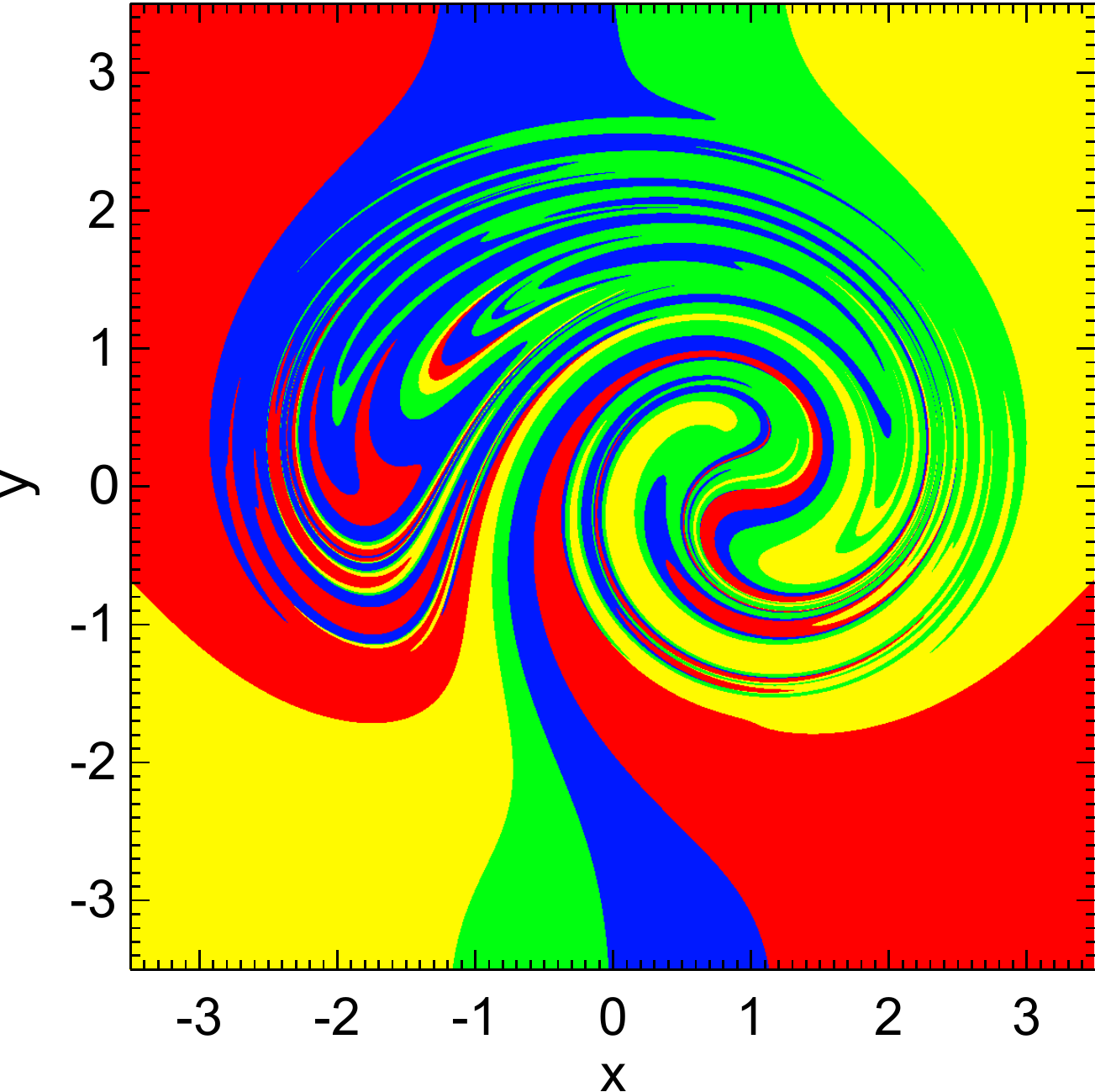}
\includegraphics[width=0.4\textwidth]{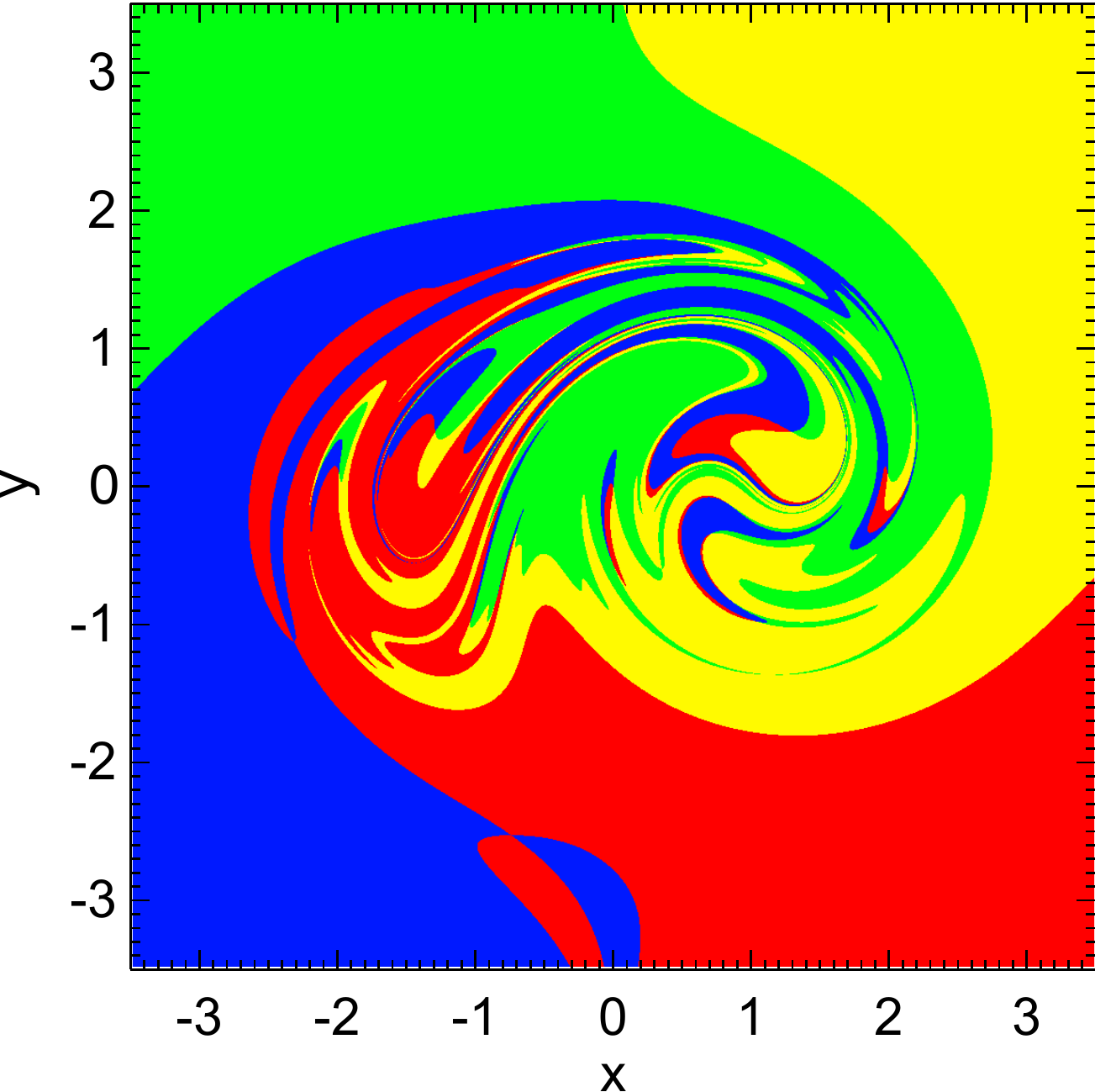}
\caption{Colour maps in the initial states of $E^{3}$ (left) and $S^{3}$ (right).}
\label{fig:colormaps}
\end{figure}

A second  way to examine field line connectivity and mixing is by way of  colour maps 
(Polymilis {\it et al.}~2003) as shown in Figure~\ref{fig:colormaps} .  These will be particularly instructive later
when considering how the systems relax to lower energy states. 
To explain how these images are generated, consider a field line threading the domain, and let the points of intersection of the field line with the lower and upper boundaries be $(x_{0},y_{0})$ and $(X,Y)$, respectively.
We make a plot over the lower boundary, colouring
the point $(x_{0},y_{0})$ red if $X>x_{0}$ and $Y>y_{0}$,  green if $X<x_{0}$ and $Y<y_{0}$, blue if $X>x_{0}$ 
and $Y<y_{0}$ and yellow if $X<x_{0}$ and $Y>y_{0}$.  
In this way each point on the boundary is coloured, excepting those periodic orbits ($X=x_{0}$, $Y=y_{0}$)
which (generically, as in these fields) lie at the intersection of all four colours.
The complex colour maps for the initial states of  $E^{3}$  and $S^{3}$  show the braided nature of the fields,
$E^{3}$ being more complex with small-scale structures filling a greater portion of the domain.

The colour map is a visual representation of the complexity in field line connectivity.  A more precise, 
formal measure, is given by a quantity known as the topological entropy. 
 Somewhat like our integrated squashing factor, this is a global measure that gives a single 
 number for the whole magnetic field.  It has the advantages both of a firm theoretical grounding 
and of being a robust quantity insensitive to  small changes in the  magnetic field. 
 There are several equivalent definitions of the topological entropy, but a convenient one is the asymptotic 
 growth rate (with $z$) of horizontal loops stretched around the magnetic field lines 
 (Newhouse \& Pignataro 1993; Thiffeault 2010). While the exact entropy depends on the full pattern 
 of magnetic field lines, a good estimate may be obtained by ensemble averaging over finite sets of field 
 lines. Applying the numerical method of Moussafir (2006), as implemented by Thiffeault (2010), and with 
 sets of 40 field lines, we find $T(E^3)\approx 3.3$ and $T(S^3)\approx 2.3$.

 {In summary, while the two magnetic fields $E^{3}$ and $S^{3}$ may be generated by  {the same amount of boundary motion}, the details of the pattern of boundary motions is crucial in determining the 
braiding pattern of the resulting fields.  From the measures presented above we see that $E^{3}$
has a significantly higher degree of complexity than $S^{3}$.}
With these differences in mind we now wish to use these two magnetic fields as initial conditions for 
resistive MHD relaxations.  We aim to show that the ability to effectively heat large regions of the loop 
depends on the nature of the photospheric driver.
Before presenting results of these experiments we first proceed to detail our numerical methods.

\section{Methods}
\label{sec:methods}

The magnetic fields corresponding to the closed-form expressions for $E^{3}$ and $S^{3}$  are not force-free.
However, the corona itself is thought to be a largely force-free environment, i.e.~one with magnetic 
fields ${\bf B}$ and associated currents $({\bf J} = (\nabla \times {\bf B})/\mu_{0}$) satisfying
$ {\bf J} \times {\bf B} \approx \mathbf{0}.$
The first stage of our experiments is therefore to use an ideal Lagrangian relaxation code 
(Craig \& Sneyd 1986) to relax the fields towards a force-free equilibrium whilst exactly preserving their topology.
In the scheme, details of which may be found in Craig \& Sneyd (1986), an artificial frictional evolution is taken to 
minimize ${\bf J} \times {\bf B}$.  The relevant output is then the final state of relaxation while the path to this state 
is not  important.

\begin{figure} 
\centering
\includegraphics[width=0.25\textwidth]{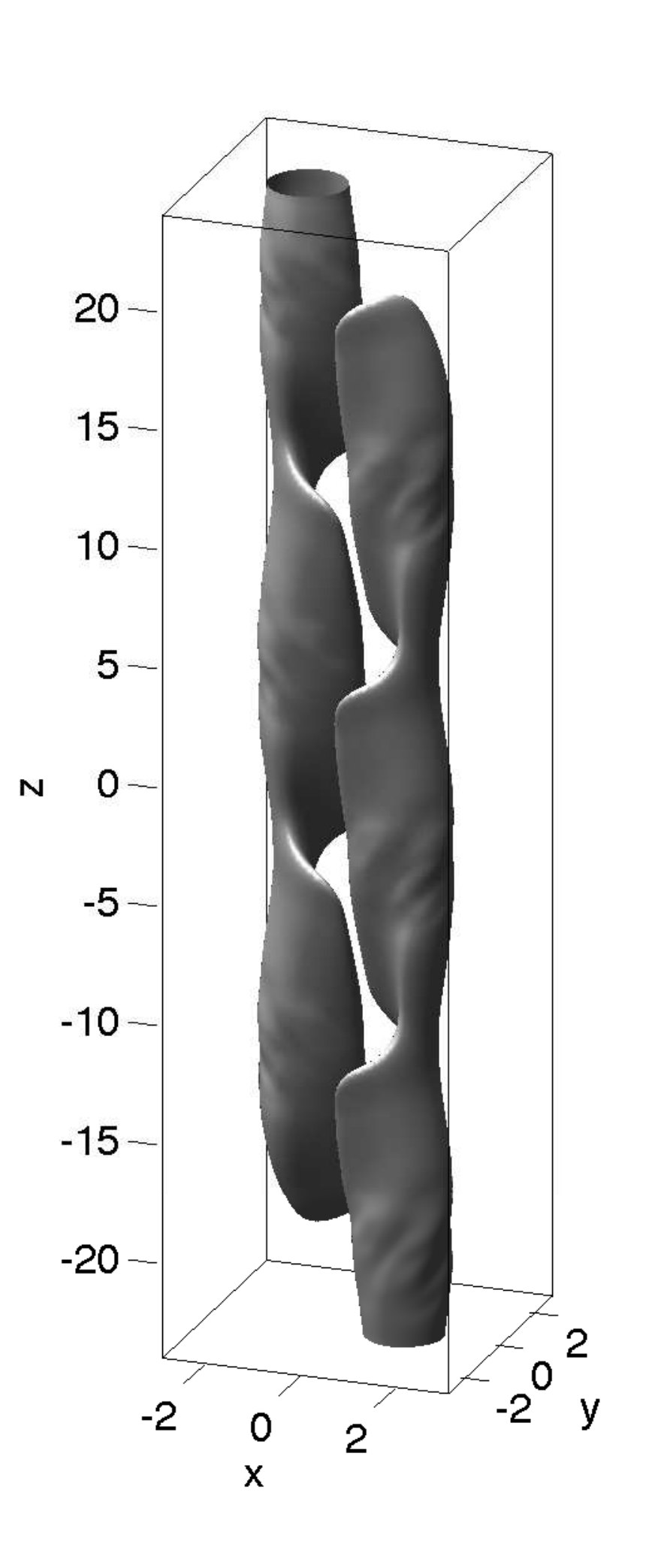}
\includegraphics[width=0.25\textwidth]{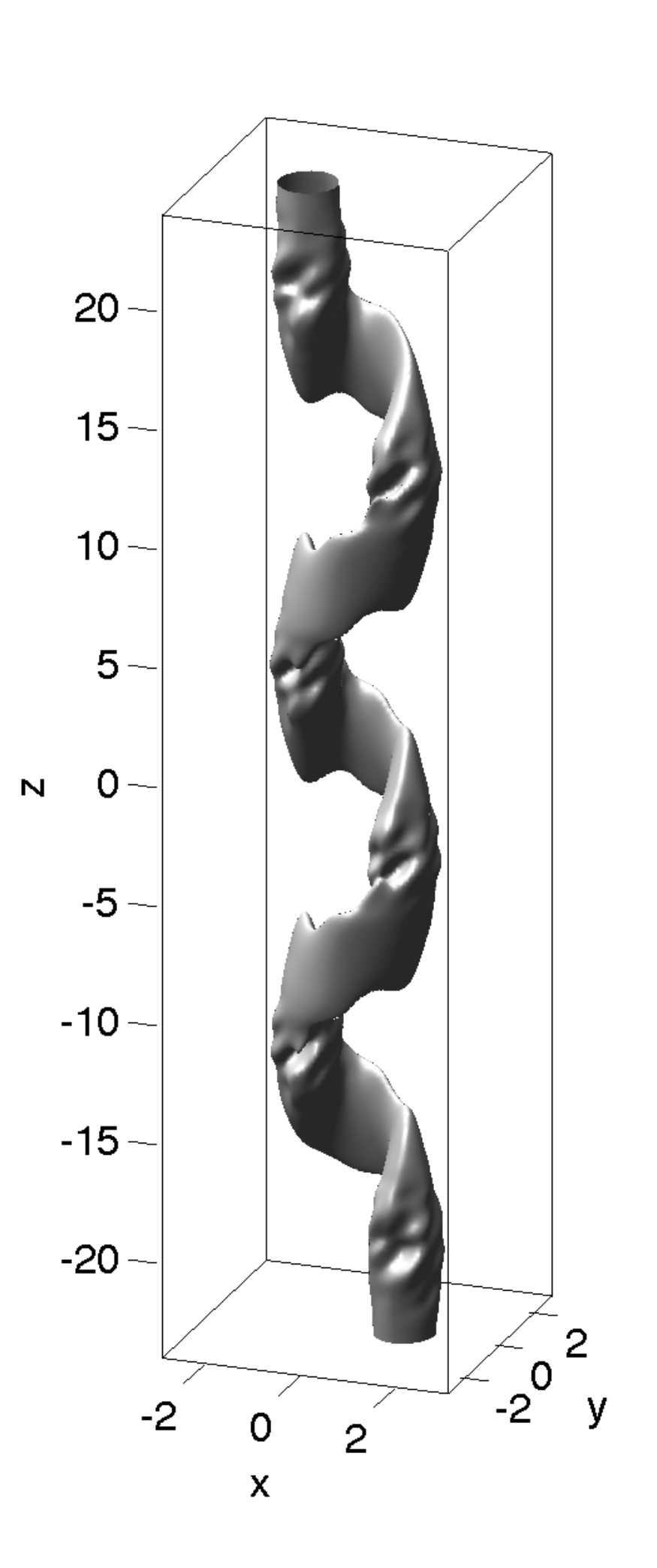} 
\caption{Isosurfaces of $\vert{\bf J}\vert$ in the initial state of the resistive MHD simulations
for $E^{3}$ (left) and $S^{3}$ (right).  The isosurfaces are taken  at 
 {25\% of the} domain maximum in both cases.} 
\label{fig:ICrmhd}
\end{figure}

 {Each of the fields $E^{3}$ and $S^{3}$ detailed in Section~\ref{sec:compare} is used as an initial condition
for this ideal relaxation, over a domain $x,y \in [-6,6]$, $z \in [-24,24]$  with a uniformly spaced grid of
$101^{3}$ points.}
The result of this procedure was described in detail for $E^{3}$ in Wilmot-Smith {\it et al.}~(2009). 
A smooth near-force--free field is obtained with large-scale
current distributions in the form of two tubes of current running through the domain
 {(see Figure~\ref{fig:ICrmhd})}. 
By `near-force--free' we mean that the maximum Lorentz force in the domain is
 $({\bf J} \times {\bf B})_{max} \approx 0.059$
  {where both $\mathcal{O}({{\bf J}})$ and $\mathcal{O}({{\bf B}}) \sim 1$}.  
 Numerical difficulties, as documented by Pontin {\it et al.} (2009)
prevent possible relaxation to a field arbitrarily close to ${\bf J} \times {\bf B} = \mathbf{0}$.
For our purposes it is enough to be close to (rather than exactly at) a force-free state since this is also 
the relevant case in the solar corona.
The  {ideal} relaxation procedure for $S^{3}$ also results in a smooth approximately force--free field
(with $({\bf J} \times {\bf B})_{max} \approx 0.057$).  A large-scale current distribution is present, now with 
one twisted current tube running through the domain  {(see Figure~\ref{fig:ICrmhd})}. 
In both cases the magnetic energy of the ideally relaxed field is reduced,  the amount of energy in 
excess of potential now being $1.286\%$ for $E^{3}$ and $1.178\%$ for $S^{3}$.
 {This will be discussed further in Section~\ref{sec:energyheating}.}

These approximately force-free fields are now used for the main body of our work
in which they are taken as initial conditions in 3D resistive MHD simulations.
In order to create the initial conditions on the regular grid required, an interpolation procedure 
must be followed.  The procedure is detailed in
Wilmot-Smith {\it et al.} ~(2010) and ensures the interpolated field remains divergence-free
to accuracies of the order of the truncation errors for sixth-order finite differences
( {$\vert \nabla \cdot {\bf B}\vert_{\textrm{max}} \approx 10^{-6}$} within the domain).
 {We use the colour-map technique to check the conservation of field line connectivity
in this interpolation step and find the conservation to be good.}

The computational setup for our MHD simulations is described below.  The experiments are conducted using the 3D non-ideal MHD code of Nordlund \& Galsgaard (1997). This is a high order finite difference code {solving} the following set of equations: \begin{eqnarray} \frac{\partial {\bf B}}{\partial t} & = & - \nabla \times {\bf E}, \label{numeq1}\\ 
{\bf E} & = & -\left( {\bf v} \times {\bf B} \right) \: + \: \eta {\bf J}, \label{numeq2}\\ 
{\bf J} & = & \nabla \times {\bf B}, \label{numeq3}\\ 
\frac{\partial \rho}{\partial t} & = & - \nabla \cdot \left( \rho {\bf v} \right), \label{numeq4}\\
\frac{\partial}{\partial t}\left( \rho {\bf v} \right) & = & - \nabla \cdot \left( \rho {\bf v} {\bf v} \: + 
        \: {\underline {\underline \tau}} \right) \: - \: \nabla P \: + \: {\bf J} \times {\bf B}, \label{numeq5}\\ 
\frac{\partial e}{\partial t} & = & -\nabla \cdot \left( e {\bf v} \right) \: - \: P \: \nabla \cdot {\bf v} \: 
           + \: Q_{visc} \: + \: Q_{J} \label{numeq6}, \end{eqnarray}
where ${\bf B}$ is the magnetic field, ${\bf E}$ the electric field, ${\bf v}$ the plasma velocity, 
$\eta$ the resistivity, ${\bf J}$ the electric current density, ${\rho}$ the density, ${\underline {\underline \tau}}$ 
the viscous stress tensor, $P$ the pressure, $e$ the internal energy, $Q_{visc}$ the viscous dissipation and 
$Q_{J}$ the Joule dissipation.  An ideal gas is assumed, and hence 
$P \: = \: \left(\gamma -1 \right) \: e \: = \: {\textstyle \frac{2}{3}}e$.
These equations have been made dimensionless by setting the magnetic permeability $\mu_0 = 1$, 
and the gas constant ($\mathcal{R}$) equal to the mean molecular weight ($M$). 
Accordingly time units are such that, for a volume with $| \rho |=| {\bf B} | = 1$,
an Alfv\'{e}n wave would travel one space unit in one unit of time. 

We solve the equations over a grid with $256^{3}$ nodes over $x,y \in [-6,6]$, $z \in [-24,24]$, though during the simulations we find the dynamics to be confined approximately within $x,y \in [-4,4]$. The magnetic field is line-tied on all boundaries throughout, and the plasma velocity is fixed to zero at these boundaries.   We obtain our initial magnetic field for the simulations via the interpolation method described above. The dimensionless plasma density is initialised as $\rho = 1$ and the thermal energy as $e=0.1$.  A spatially uniform resistivity model is taken, with $\eta = 0.001$ for both simulations.  

In the energetic considerations of Section~\ref{sec:energyheating} dimensional quantities are recovered in order to  {demonstrate the implications of our results for the solar corona}. In order to do this three characteristic values 
(here $B_{0}$, $l_{0}$ and $\rho_{0}$) should be chosen and the following relations taken:
\[ v_{0} = \frac{l_{0}}{t_{0}}, \ B_{0} = v_{0} \sqrt{\mu_{0} \rho_{0}}, \
J_{0} = \frac{B_{0}}{\mu_{0}l_{0}}, \ E_{0} = v_{0}B_{0}, \]
\[ e_{0} = \rho_{0} v_{0}^{2}, \ T_{0} = \frac{{\bar \mu} v_{0}^{2}}{R},\]
with  $\mu_{0} = 4\pi \times 10^{-7} \textrm{ H}  \textrm{ m}^{-1}$, 
${\bar \mu} = 0.6$ and $R = 8.3 \times 10^{3}  \textrm{ m}^{2}  \textrm{ s}^{-2}  \textrm{ K}^{-1}$
(we use mks units throughout). For clarity we initially present results in the dimensionless units to allow the 
reader to adjust the chosen solar parameters as desired.

Isosurfaces of current in the initial states for the resistive MHD simulations of $E^{3}$ and $S^{3}$ 
are shown in Figure~\ref{fig:ICrmhd}. The current isosurfaces are at $\vert {\bf J} \vert = 0.441 $ for $E^{3}$ and 
 $\vert {\bf J} \vert =0.489$ for $S^{3}$. These values may be compared with typical magnetic field 
 strengths of $\vert {\bf B} \vert =1$ showing that, in both cases, these current ribbons are weak.

\section{Results}
\label{sec:results}

\begin{figure*}[tbp] 
\centering
\includegraphics[width=0.19\textwidth]{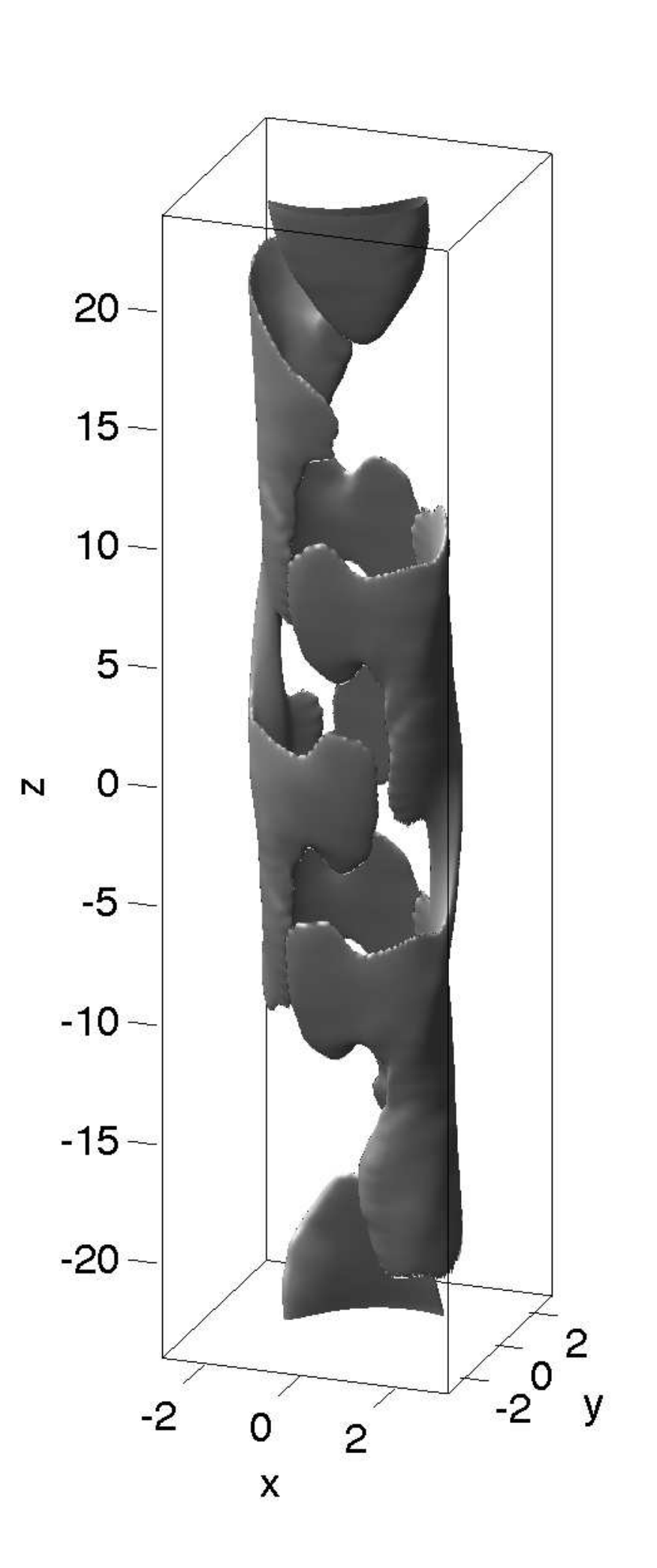} 
\includegraphics[width=0.19\textwidth]{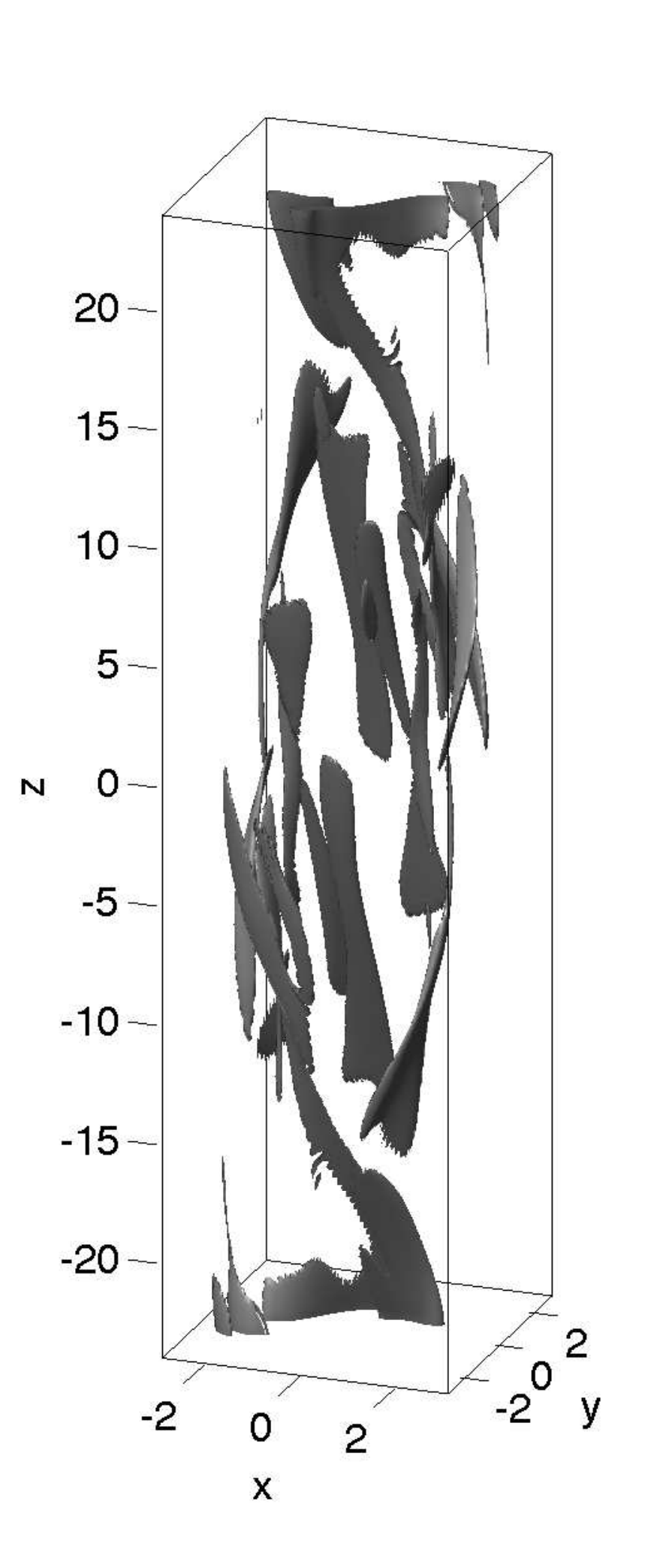} 
\includegraphics[width=0.19\textwidth]{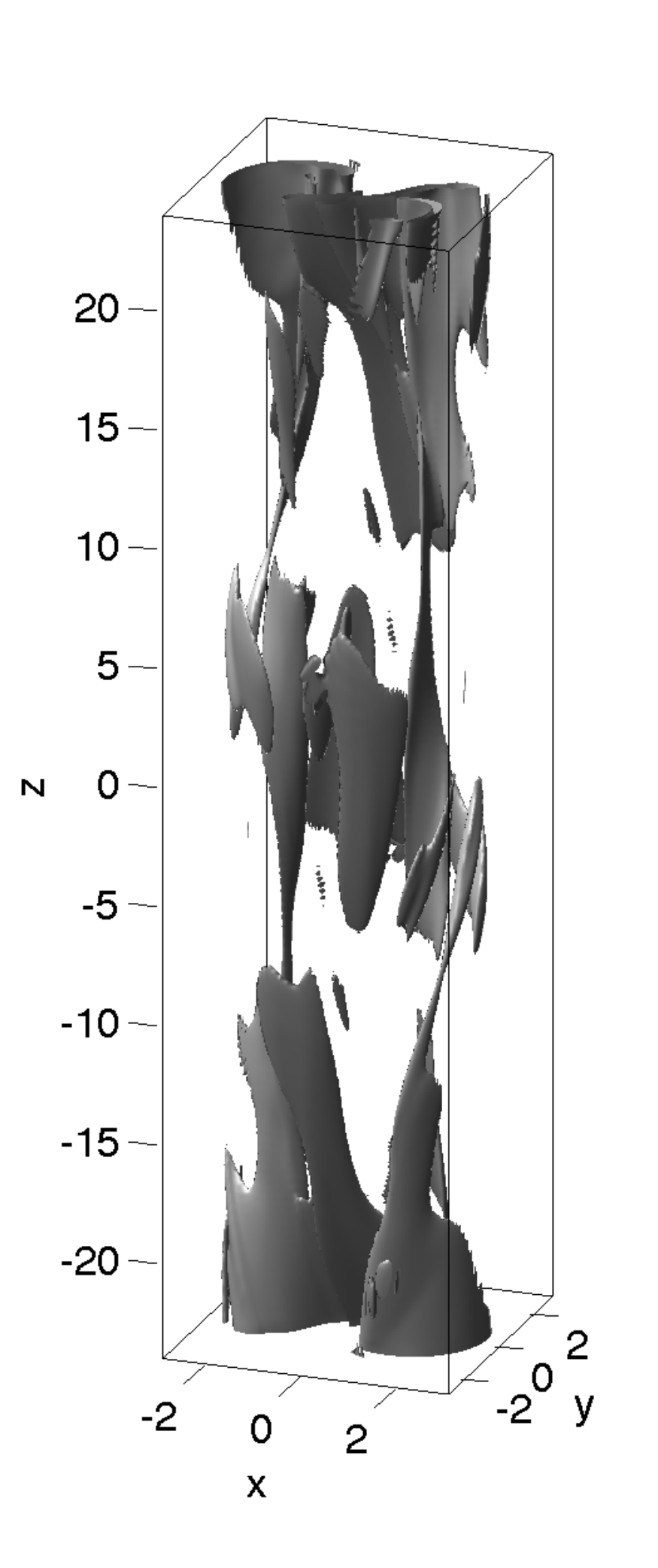} 
\includegraphics[width=0.19\textwidth]{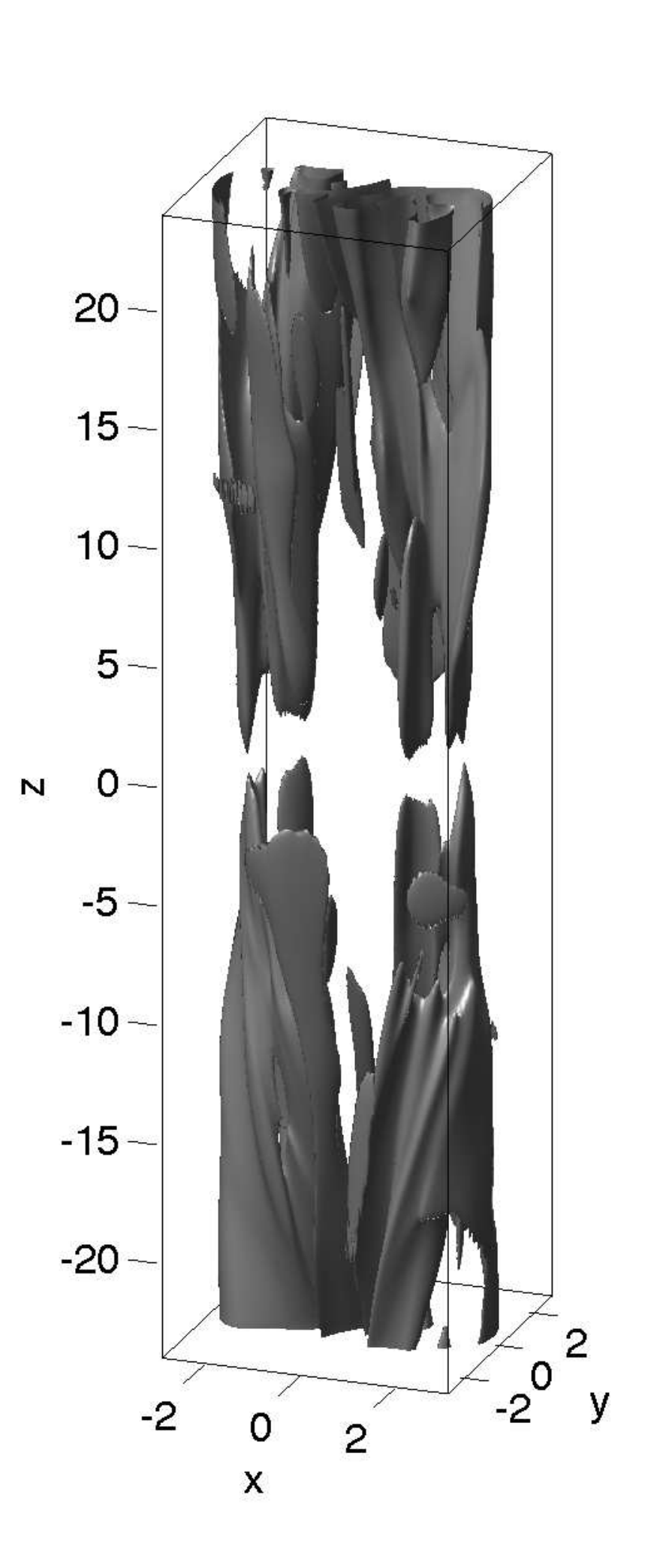} 
 \includegraphics[width=0.19\textwidth]{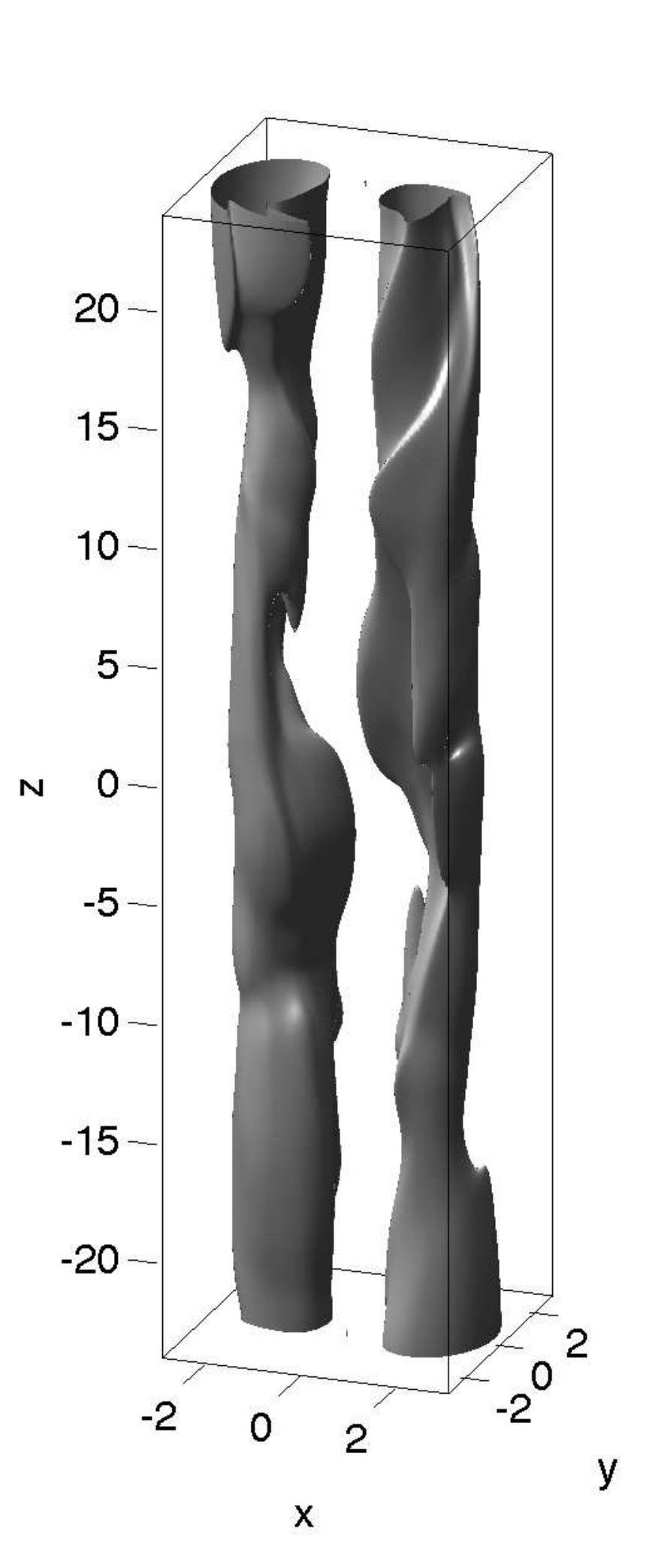} 
 
\includegraphics[width=0.19\textwidth]{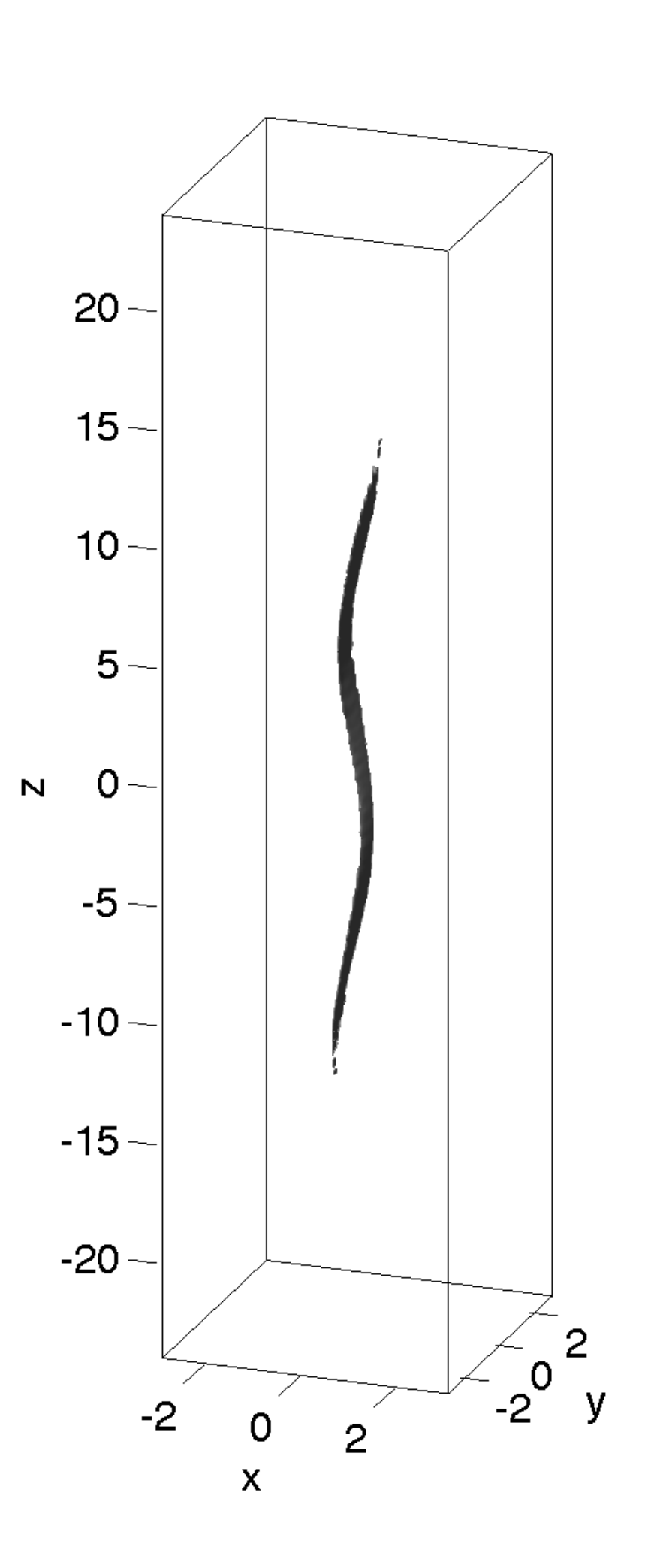} 
\includegraphics[width=0.19\textwidth]{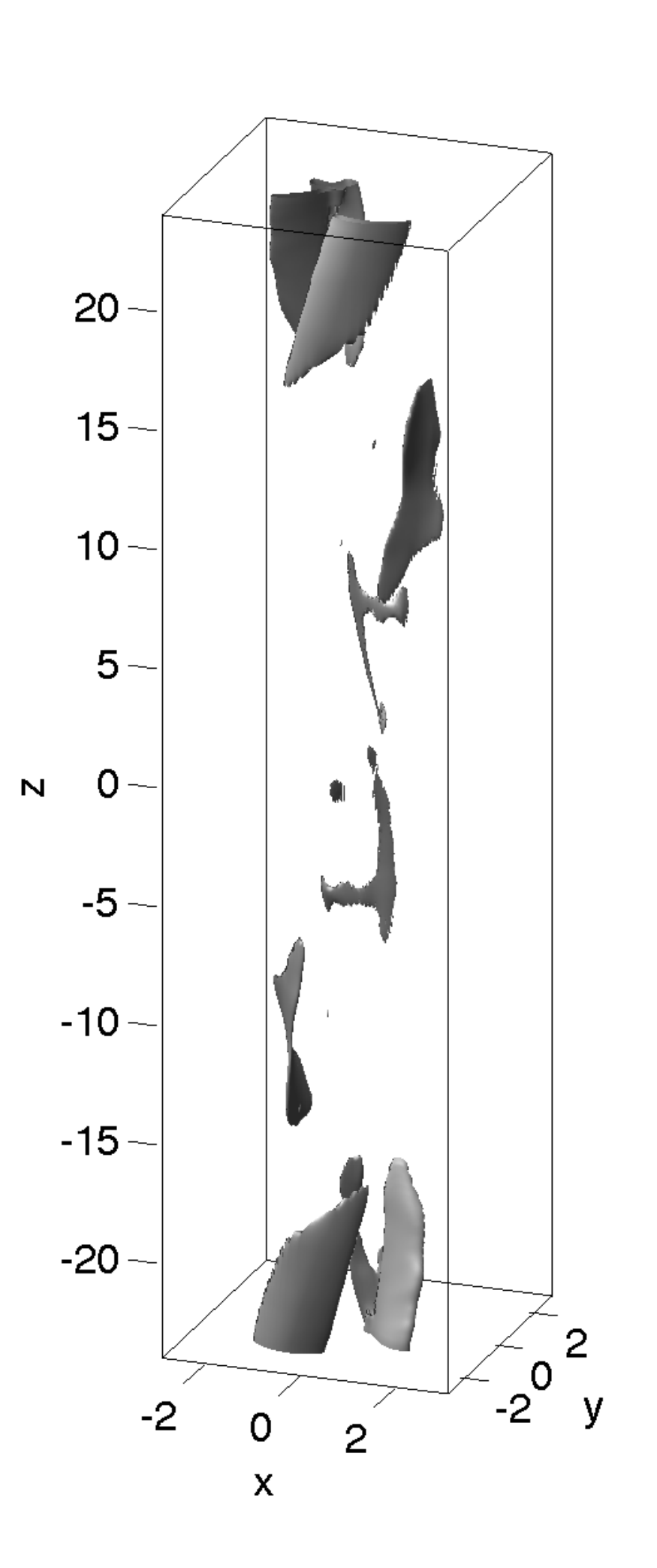} 
\includegraphics[width=0.19\textwidth]{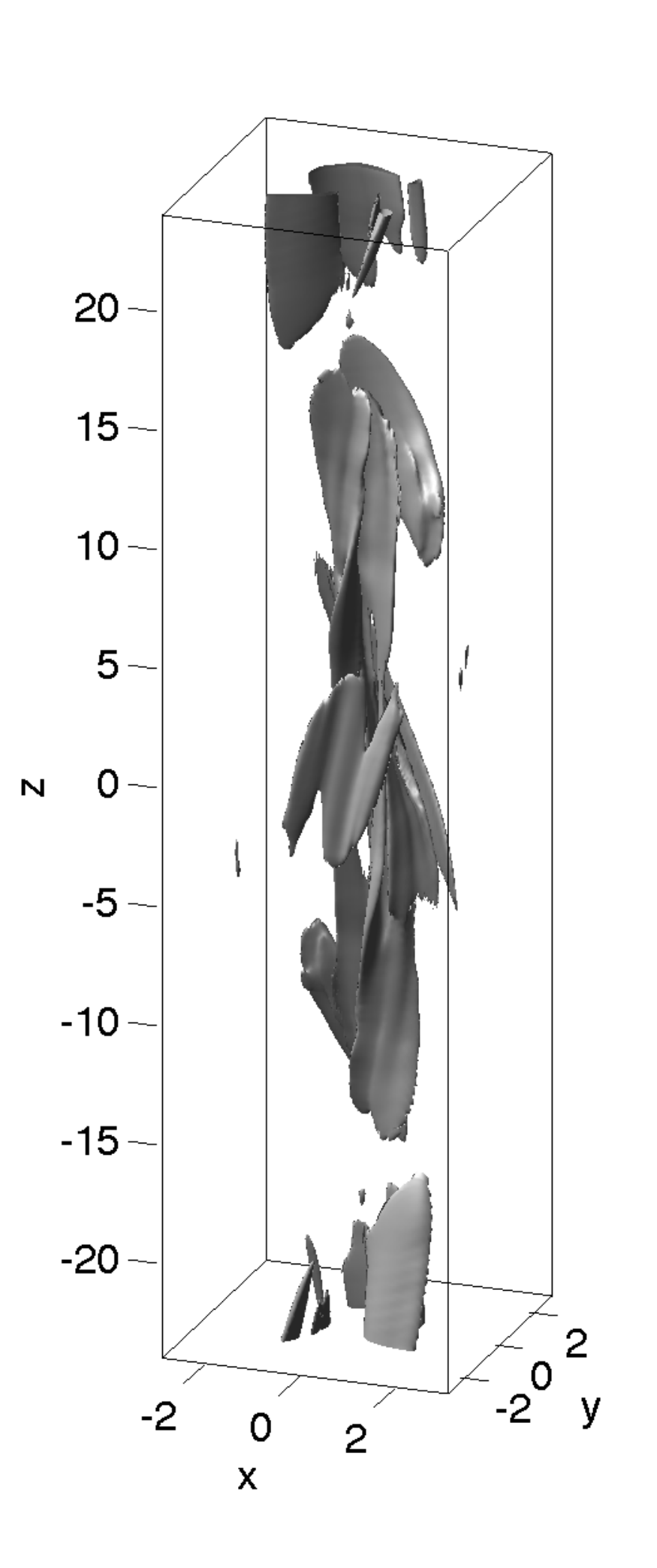} 
\includegraphics[width=0.19\textwidth]{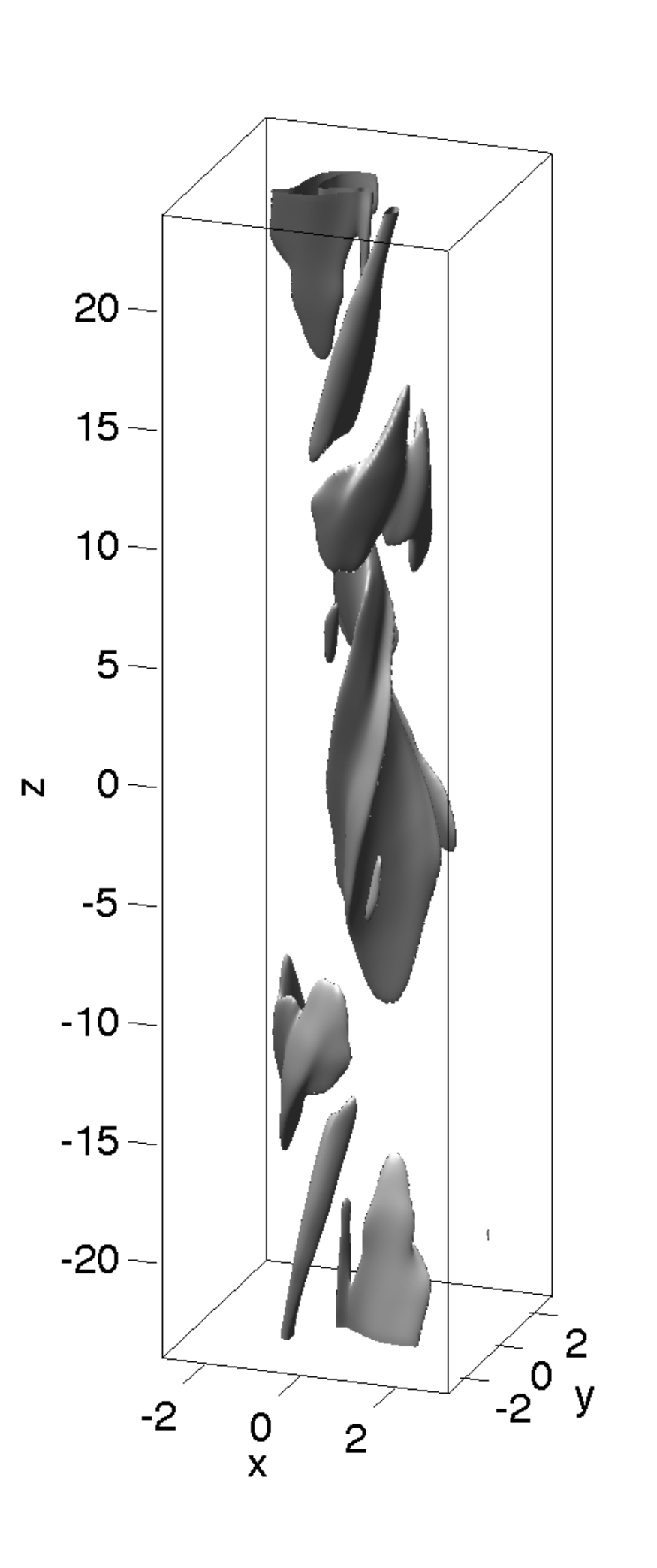} 
 \includegraphics[width=0.19\textwidth]{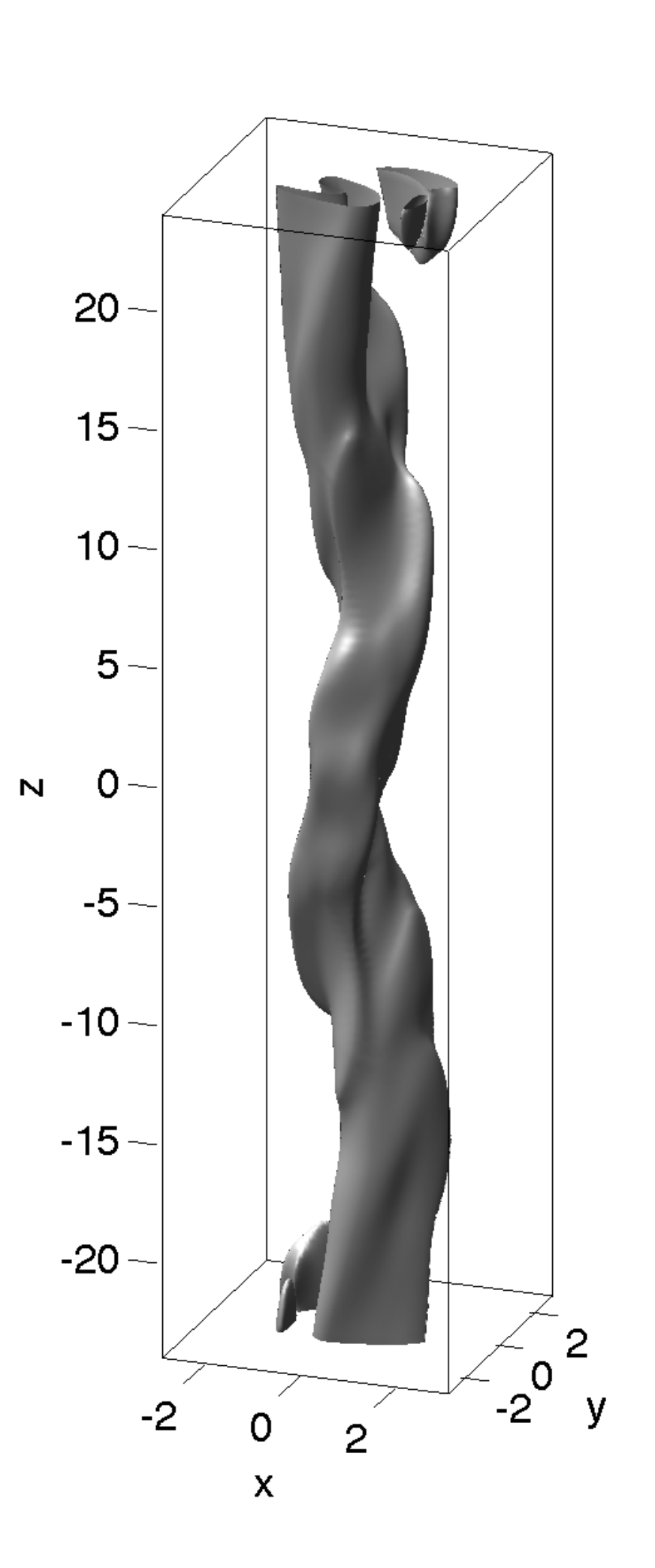} 
\caption{Isosurfaces of current at $50\%$ of the domain maximum for 
$E^{3}$ (upper panel) and $S^{3}$ (lower panel) at times $t=10$, $50$, $80$, $120$ and $350$.}
\label{fig:JisoS}
\end{figure*}

The basic  {properties of the resistive relaxation of}
 $E^{3}$ have been described in Wilmot-Smith {\it et al.}~(2010) and 
Pontin {\it et al.}~(2011).  The braided field is found to be unstable (although the precise nature of the
instability is yet to be determined) and in the early stages of the resistive evolution two main thin current layers 
are formed in the central regions of the domain.   
 {These fragment into an increasingly complex pattern of current layers
as the relaxation proceeds.}
The lower energy end-state of the relaxation is found to consist of two unlinked
twisted flux tubes of opposite twist.

A  resistive relaxation is also found to take place for the field $S^{3}$.  In this paper we aim to 
compare the two relaxation events in order to determine how the initial field configuration affects the quantity
and spatial distribution of energy released and the final equilibrium state reached.
We begin by discussing the basic nature of the relaxation events and final states (Section~\ref{sec:topo}) 
before discussing findings related to energy and heating in Section~\ref{sec:energyheating}.

\subsection{Topological properties}
\label{sec:topo}

\begin{figure}[htbp] 
\centering
\includegraphics[height=0.185\textheight]{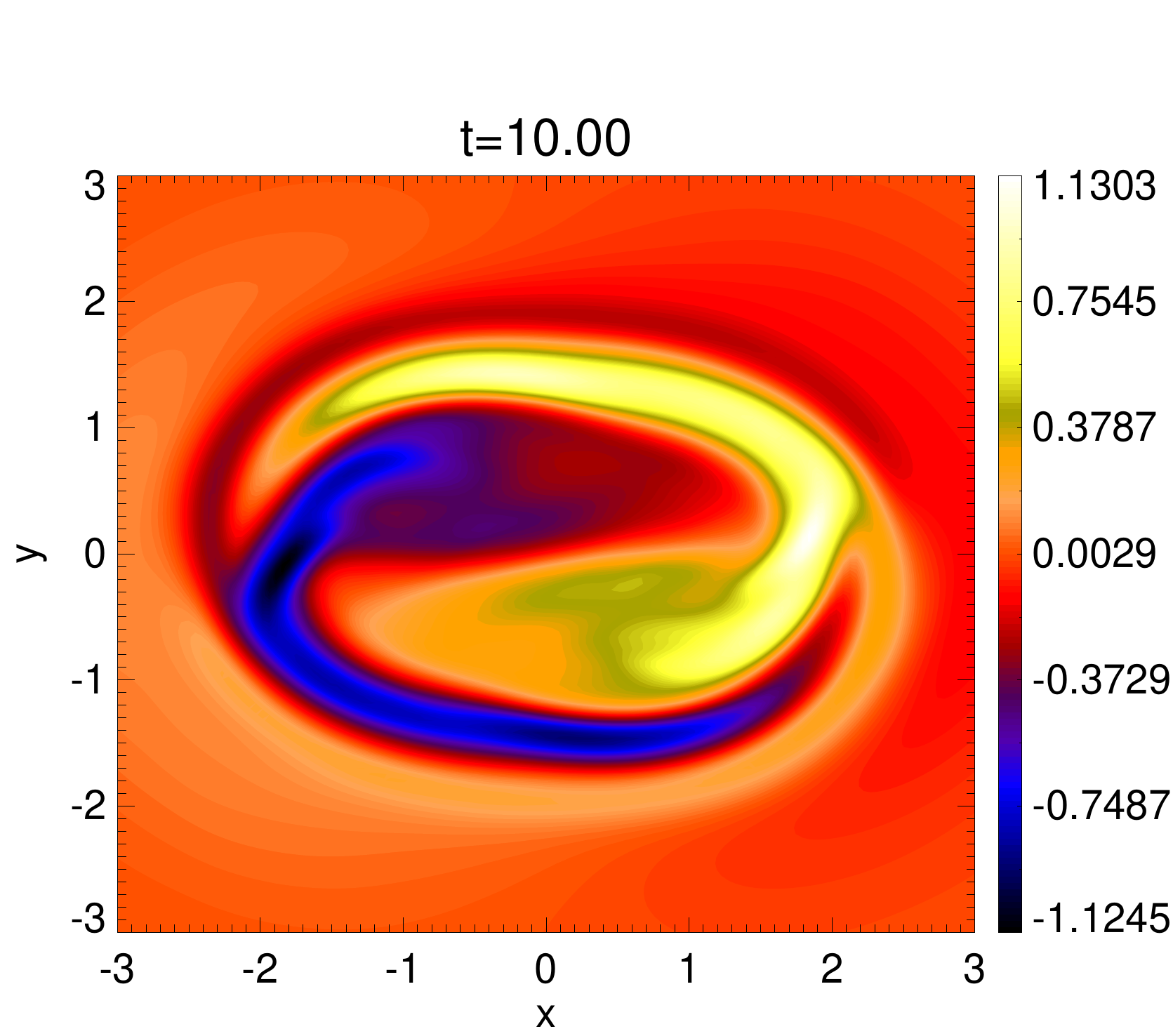} 
\includegraphics[height=0.185\textheight]{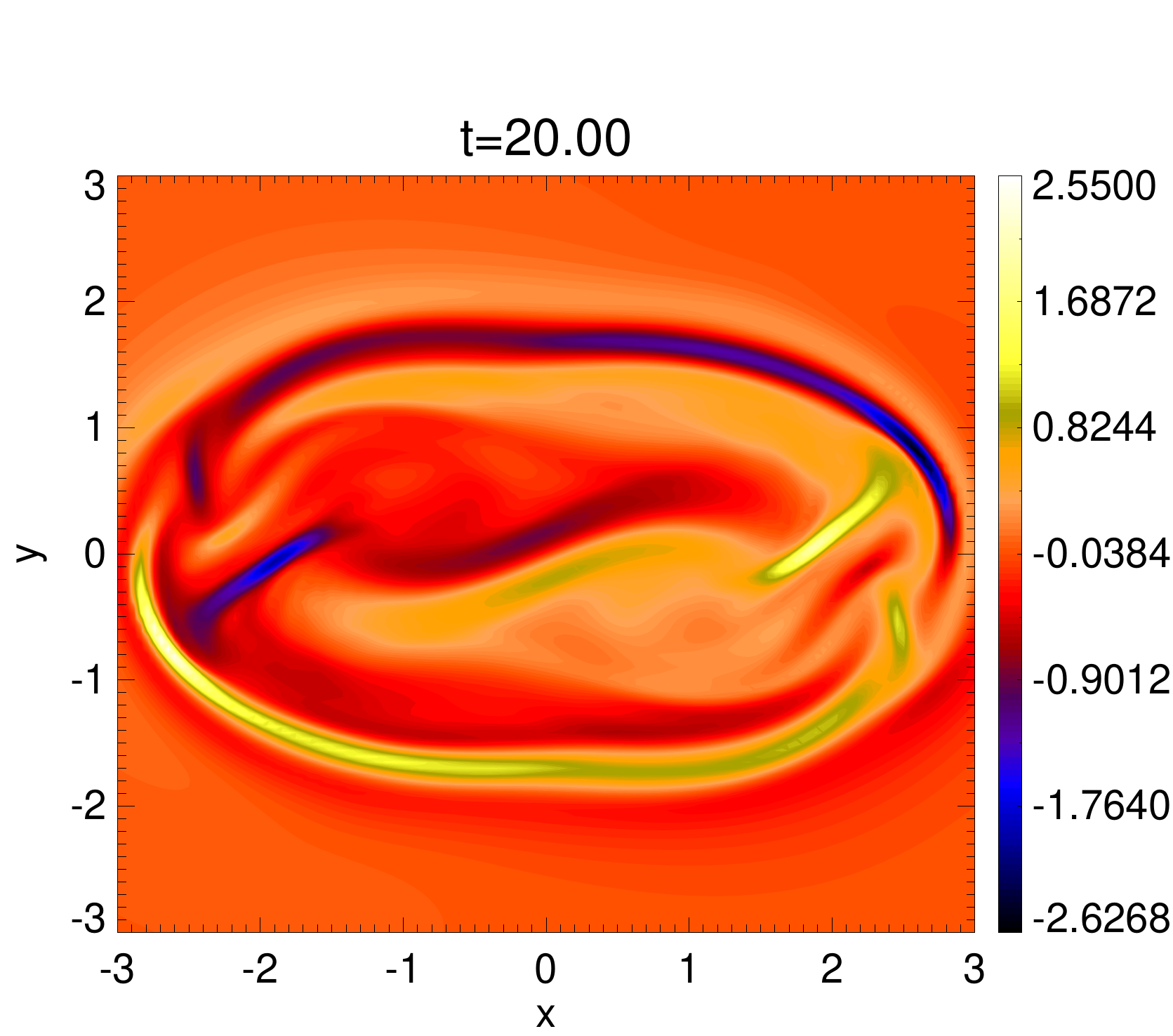} 
\includegraphics[height=0.185\textheight]{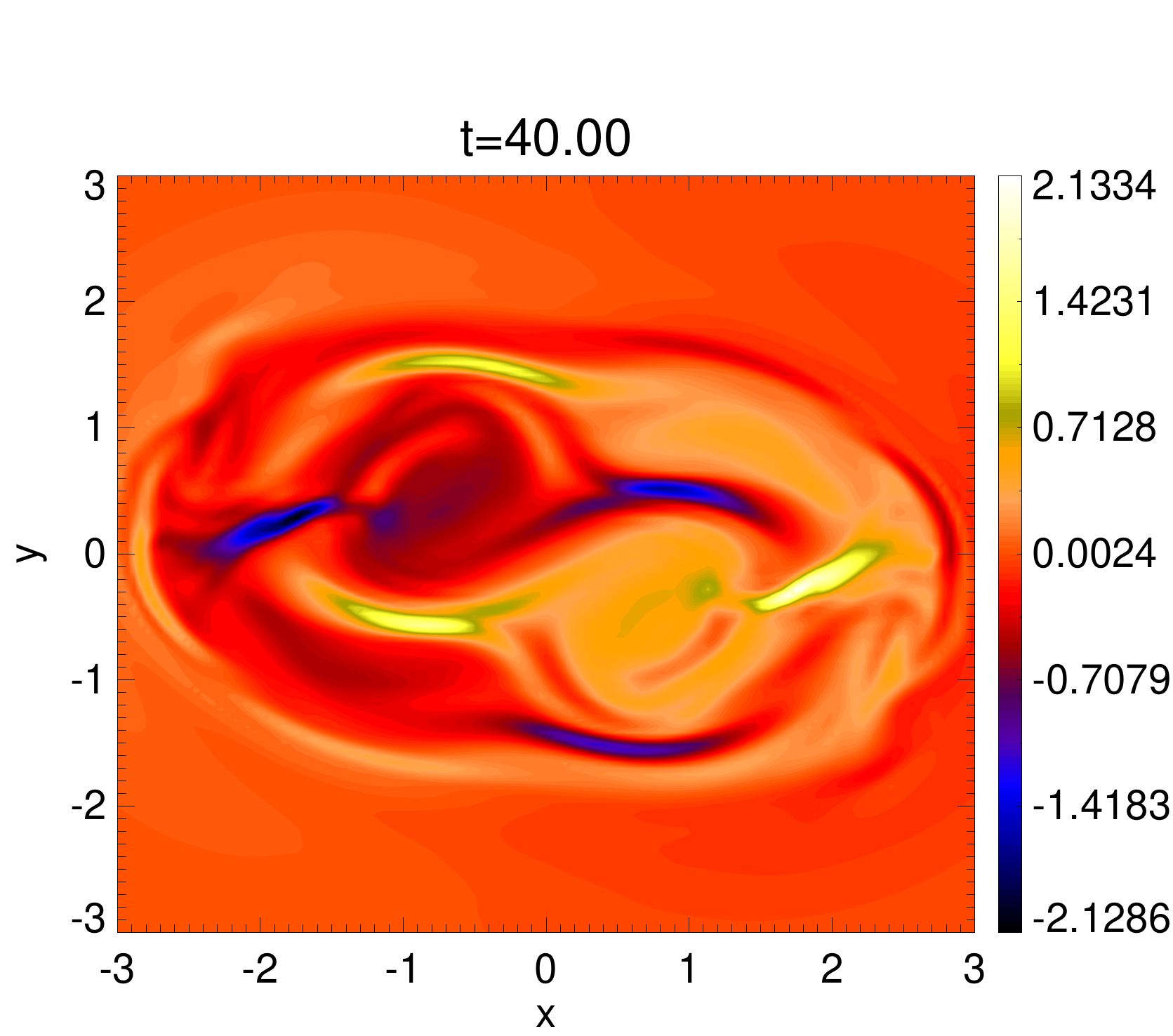} 
\includegraphics[height=0.185\textheight]{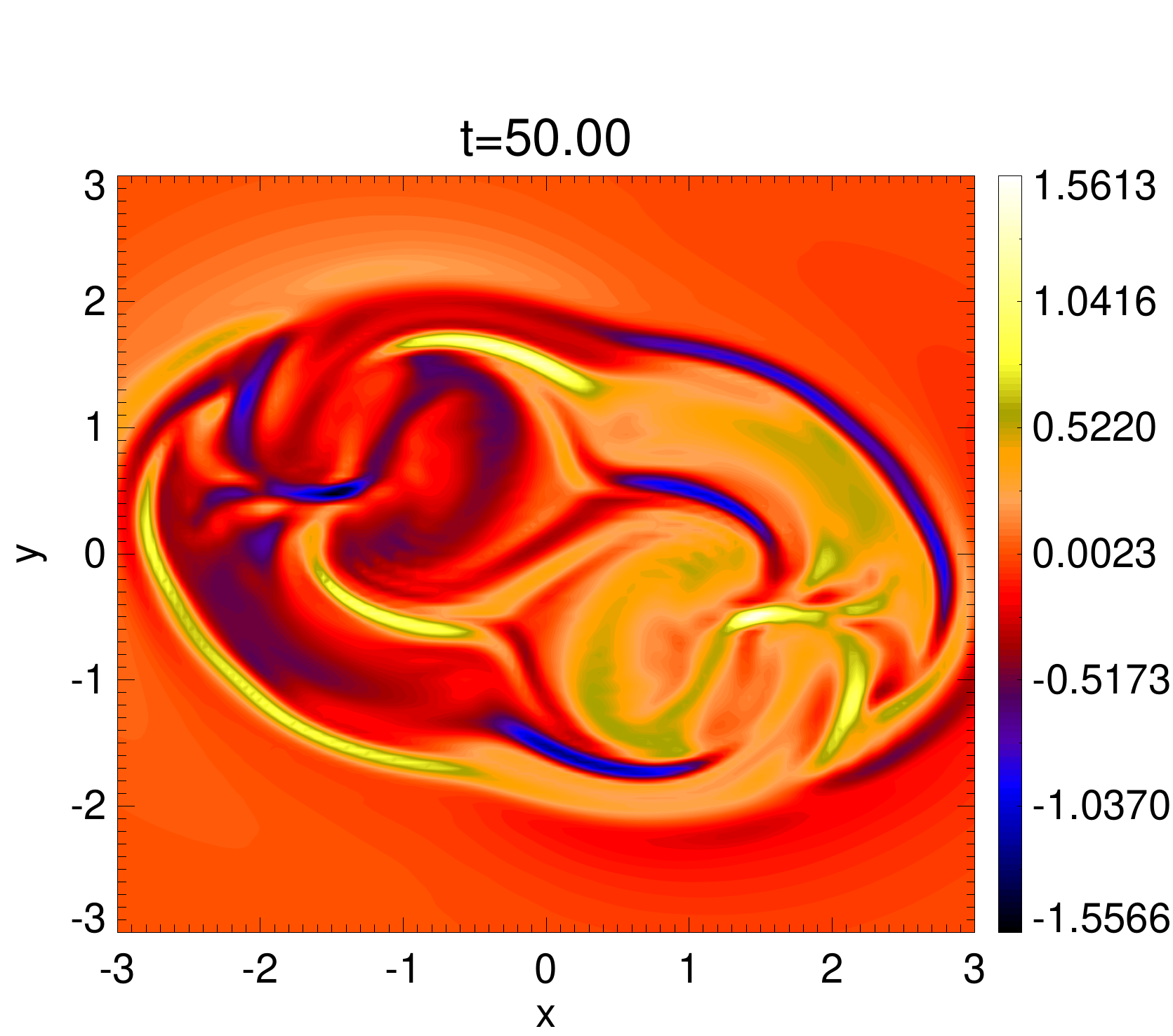} 
\includegraphics[height=0.185\textheight]{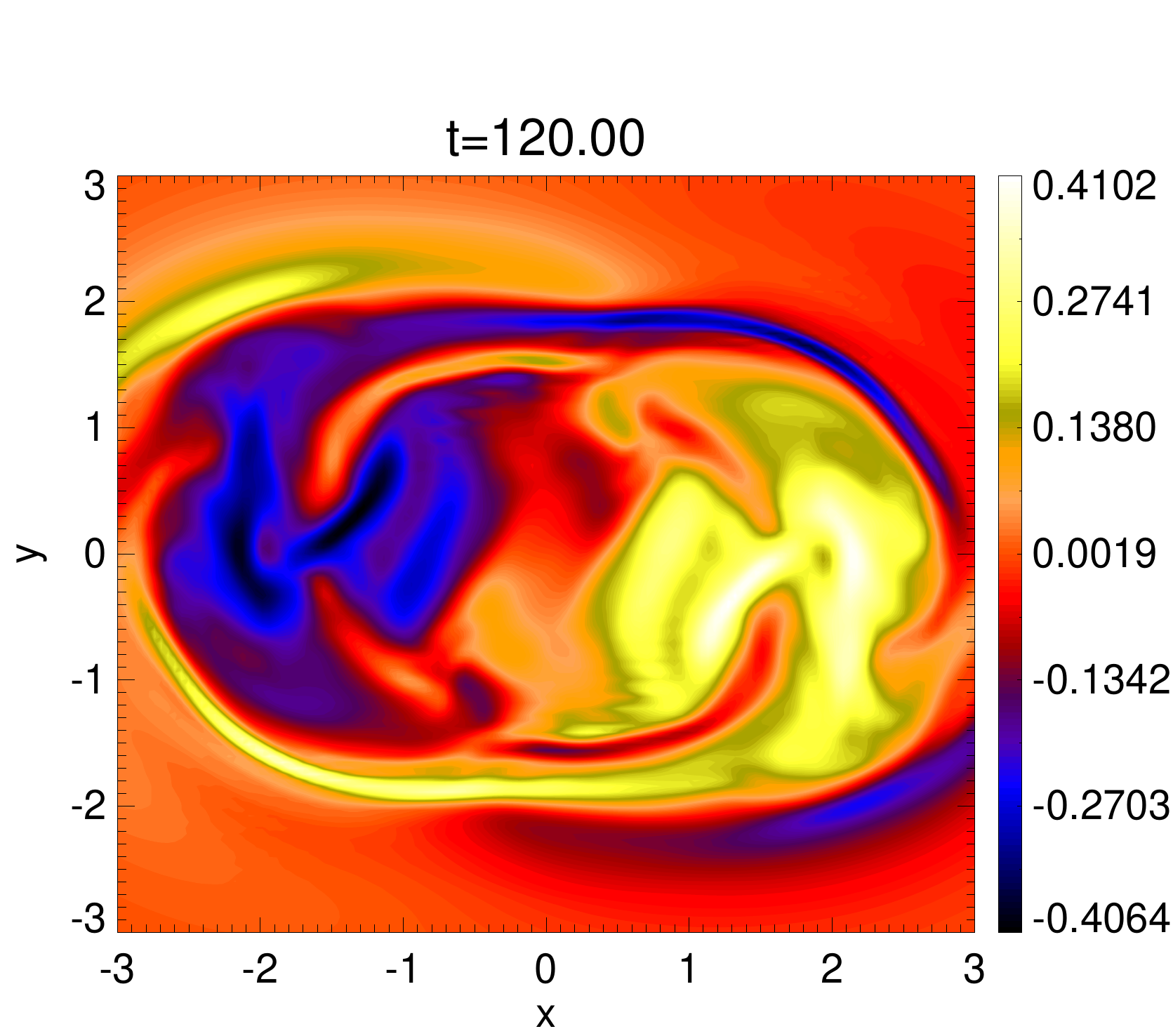} 
\includegraphics[height=0.185\textheight]{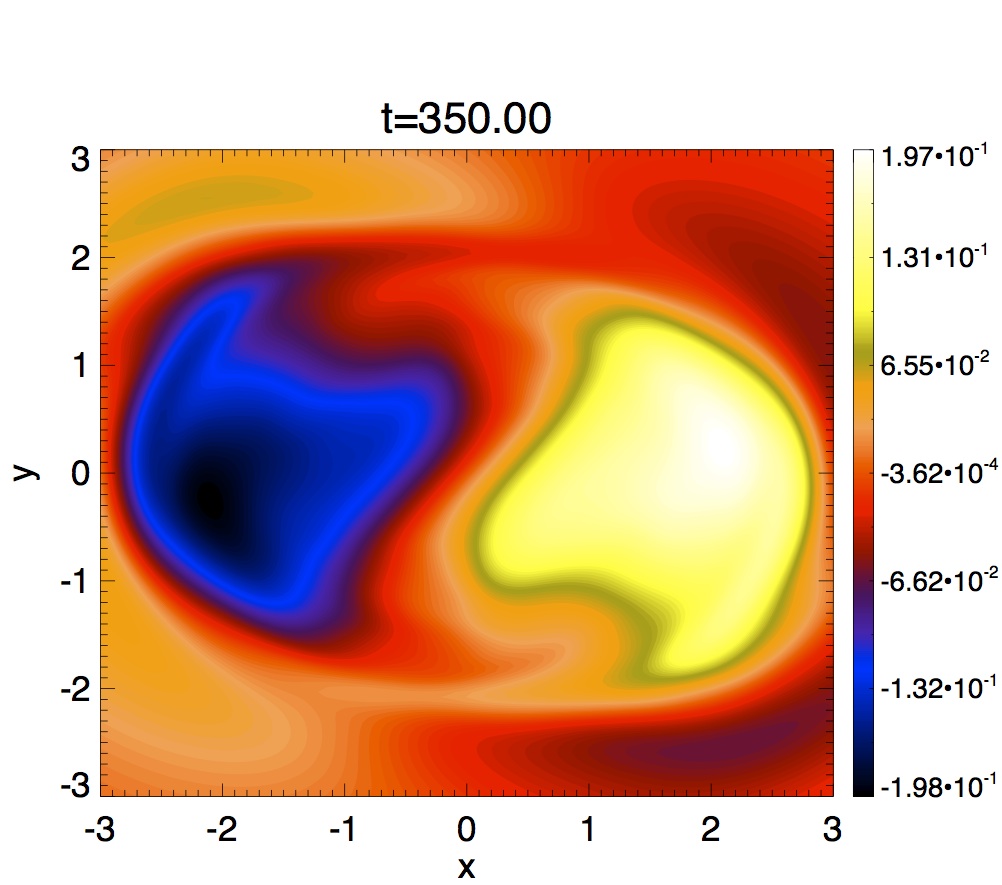}
\caption{Contours of the vertical $({\bf e}_{z})$ component of current in the $z=0$ plane
for $E^{3}$ at times as indicated in each individual image.}
\label{fig:JcontoursE}
\end{figure}

\begin{figure}[htbp] 
\centering
\includegraphics[height=0.185\textheight]{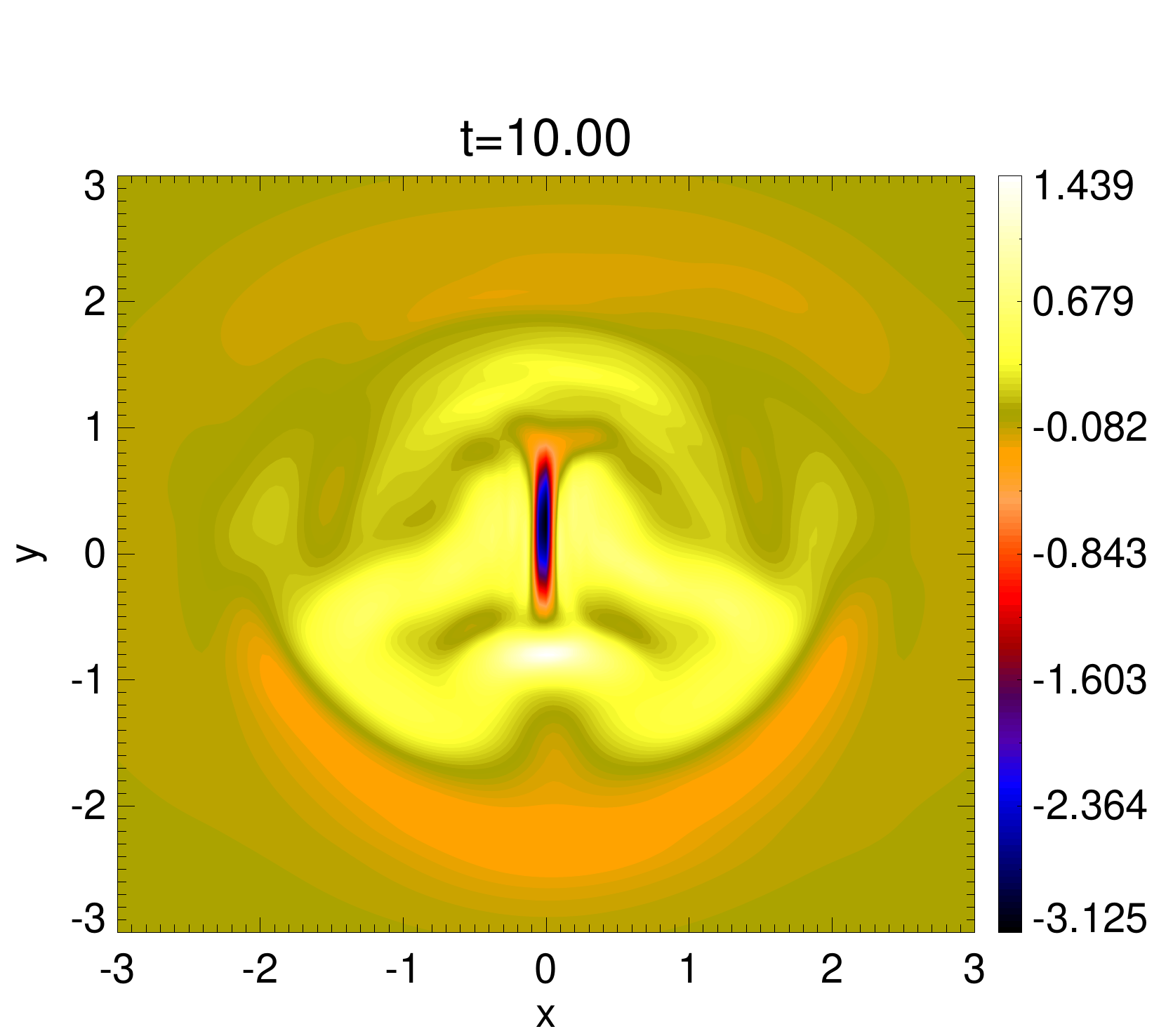} 
\includegraphics[height=0.185\textheight]{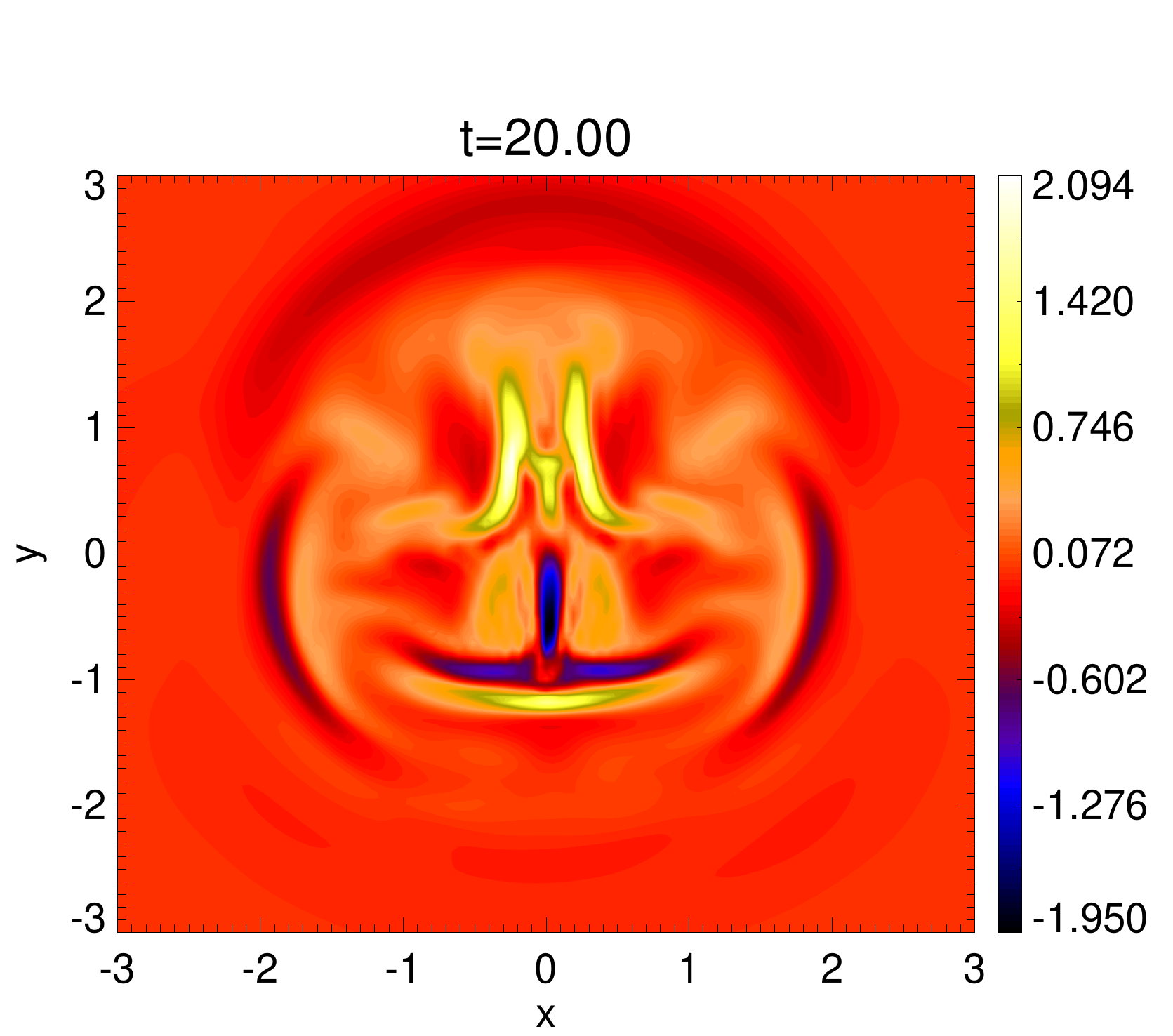} 
\includegraphics[height=0.185\textheight]{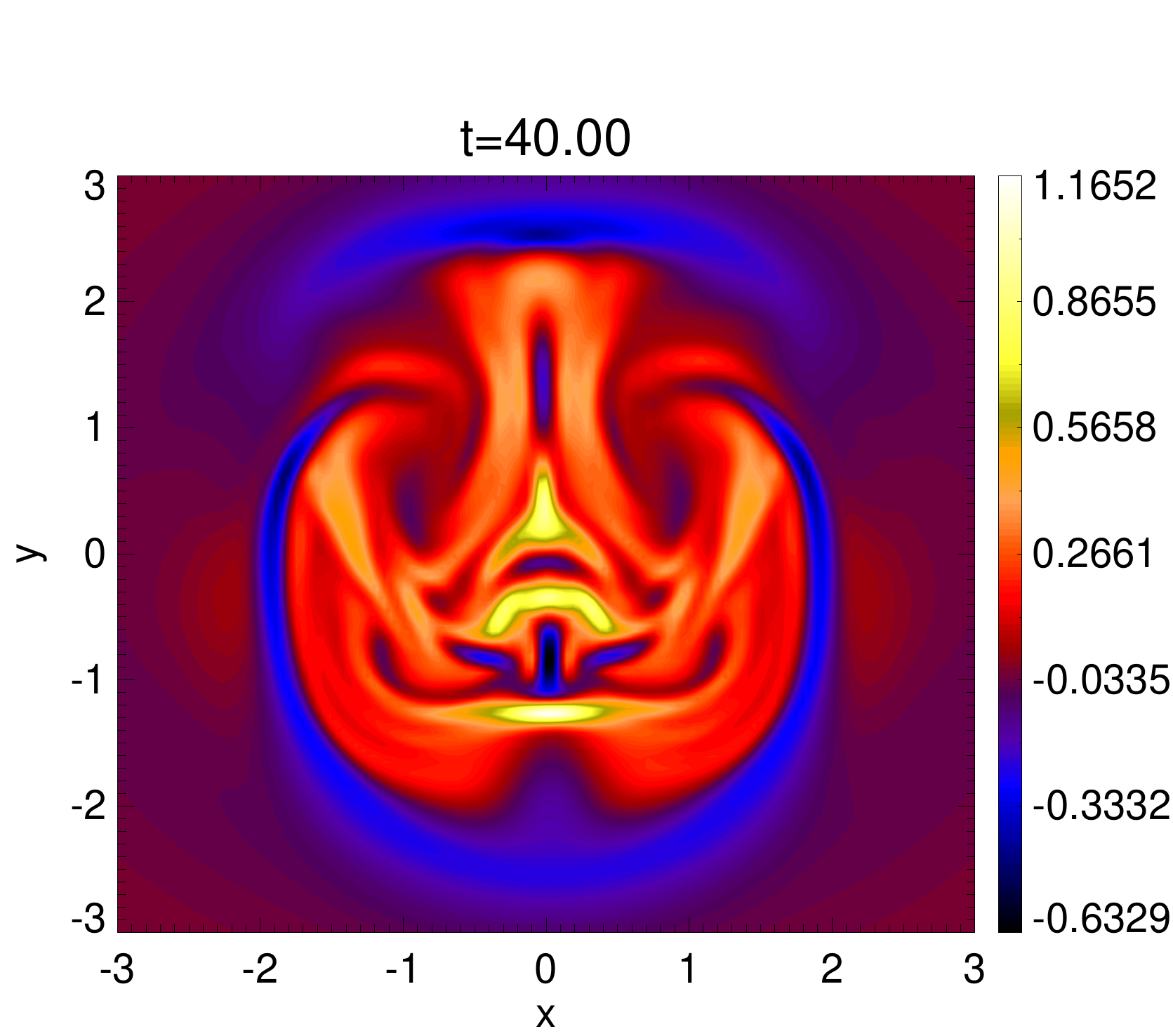} 
\includegraphics[height=0.185\textheight]{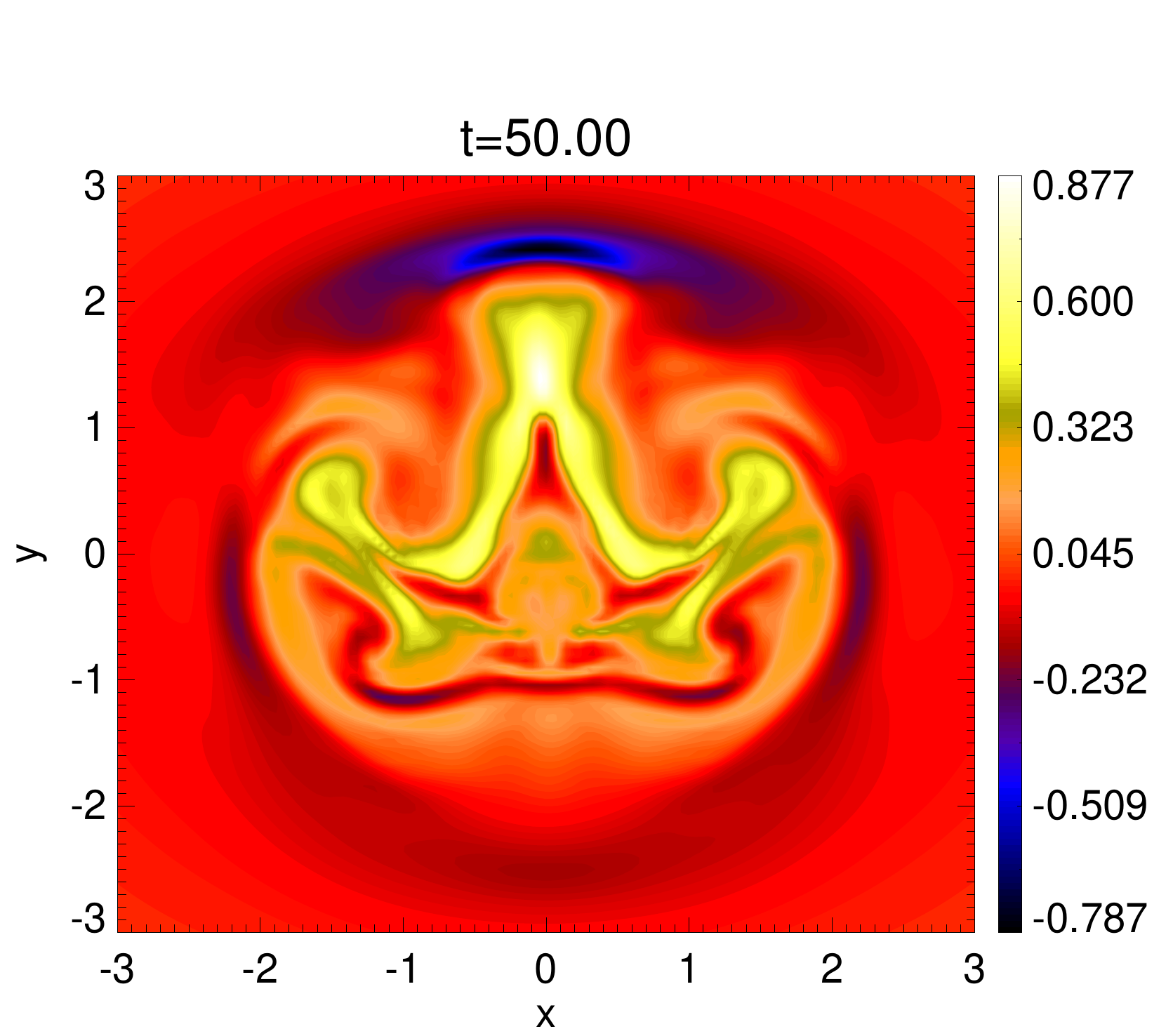} 
\includegraphics[height=0.185\textheight]{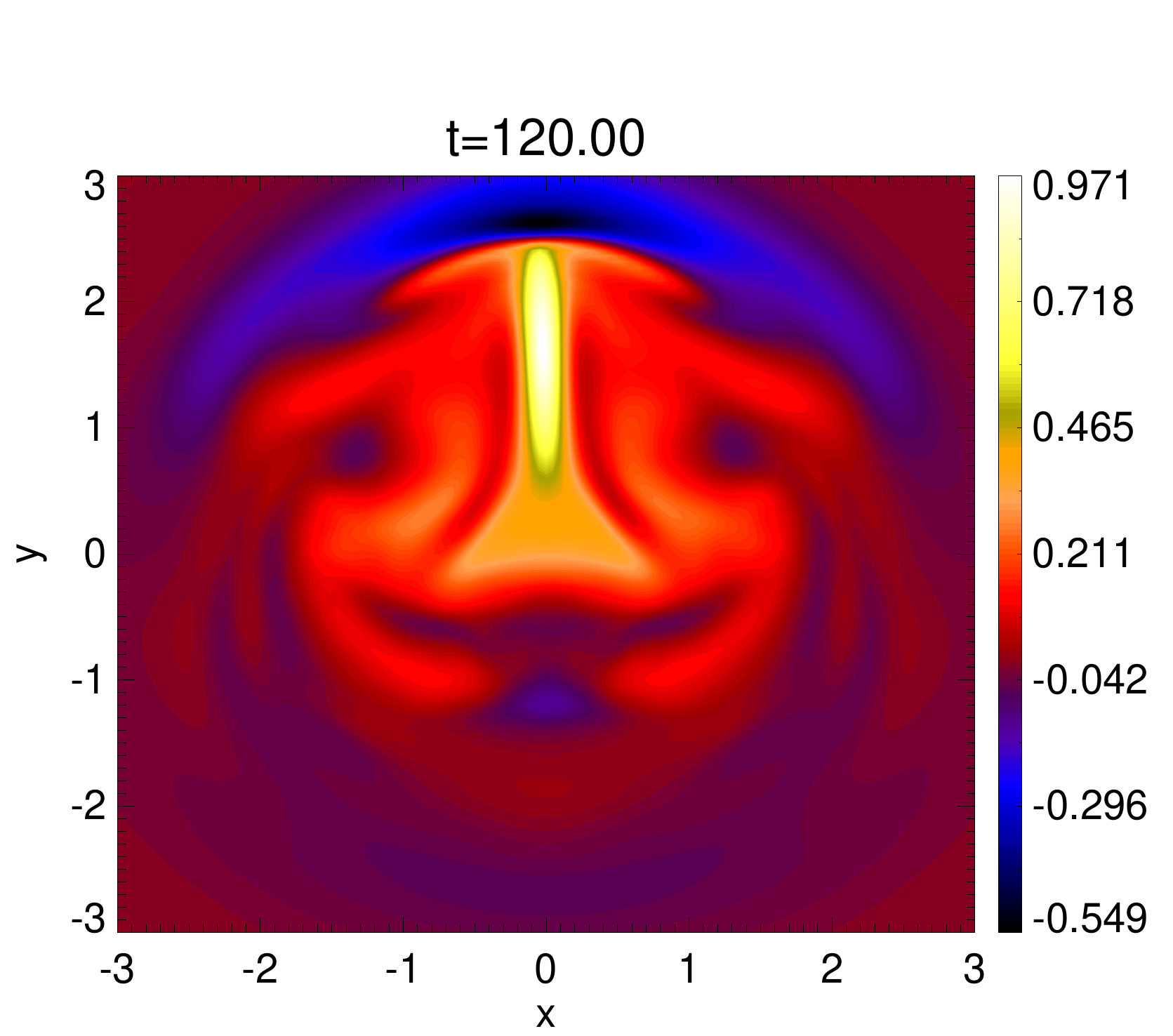} 
\includegraphics[height=0.185\textheight]{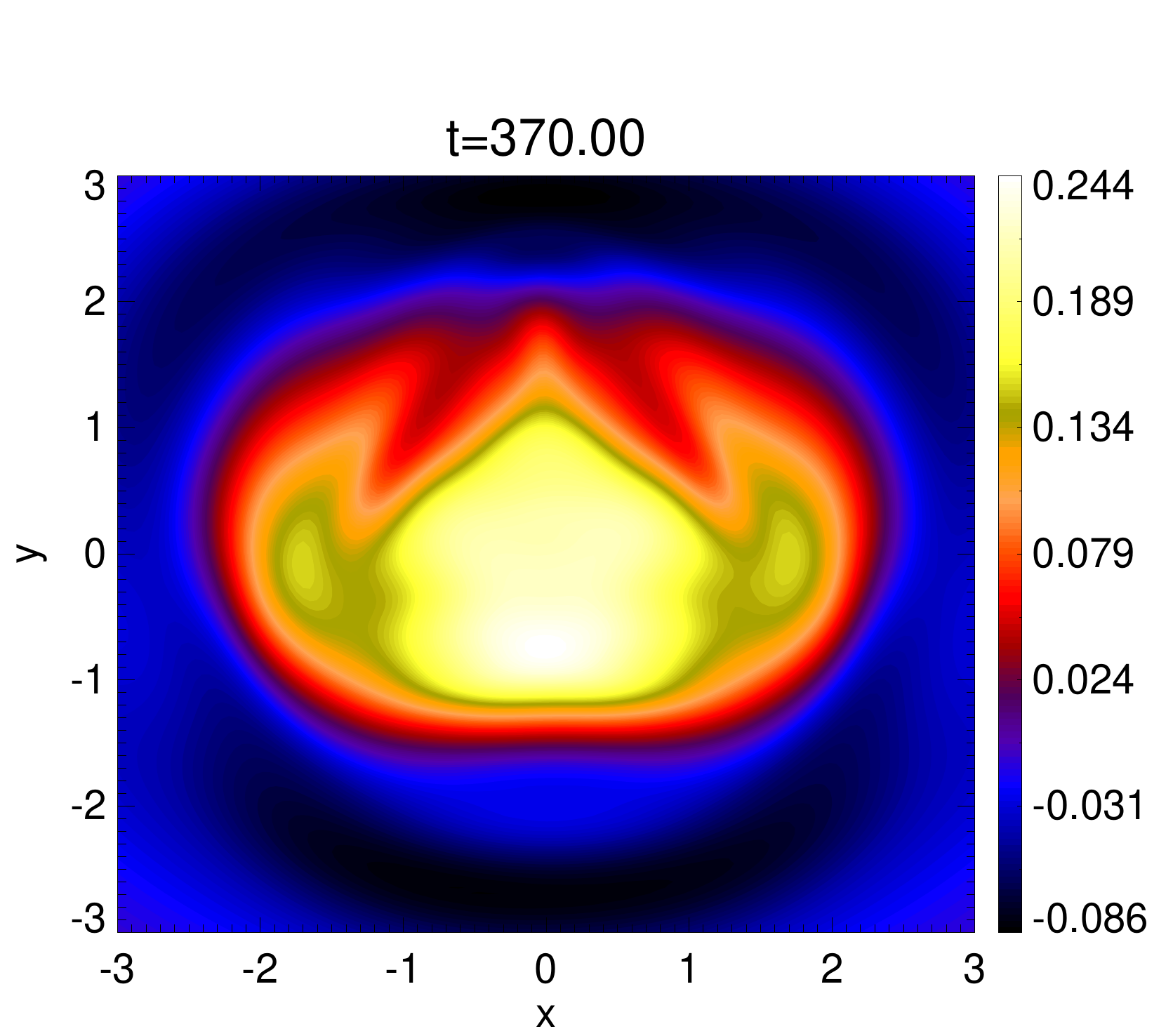}
\caption{Contours of the vertical $({\bf e}_{z})$ component of current in the $z=0$ plane
for $S^{3}$ at times as indicated in each individual image.}
\label{fig:JcontoursS}
\end{figure}

\begin{figure}[htbp] 
\centering
\includegraphics[width=0.52\textwidth]{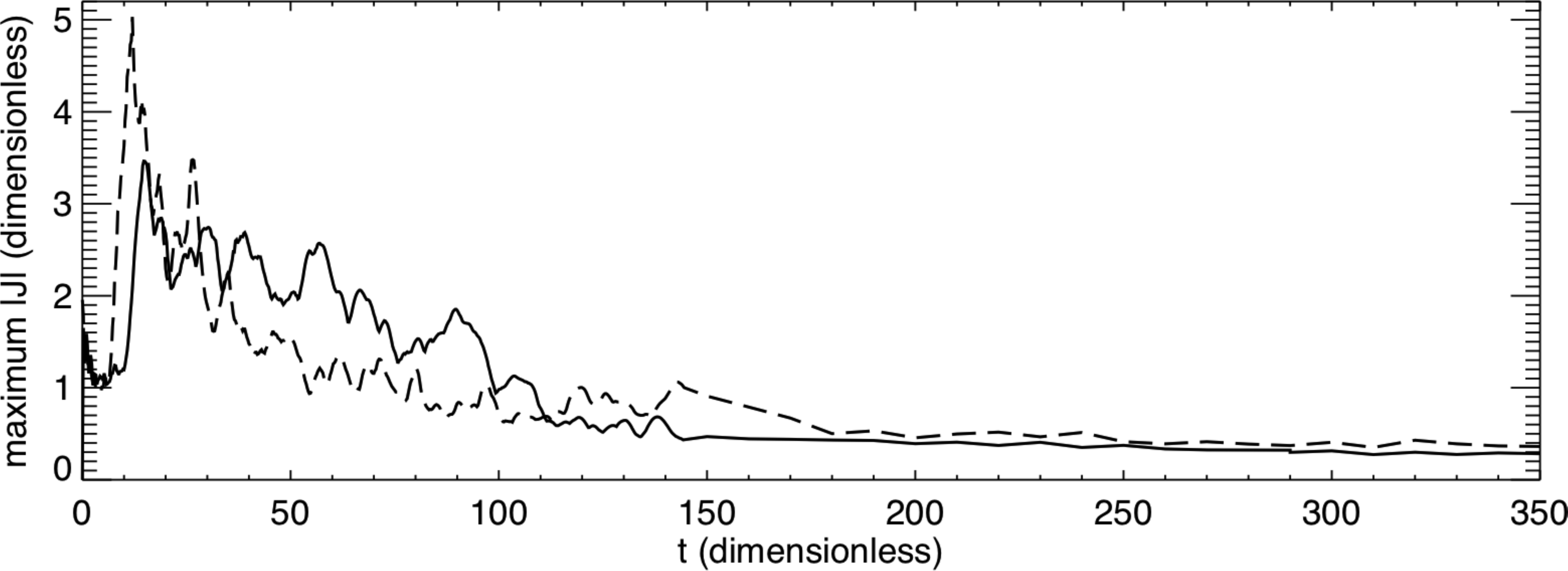}
\includegraphics[width=0.52\textwidth]{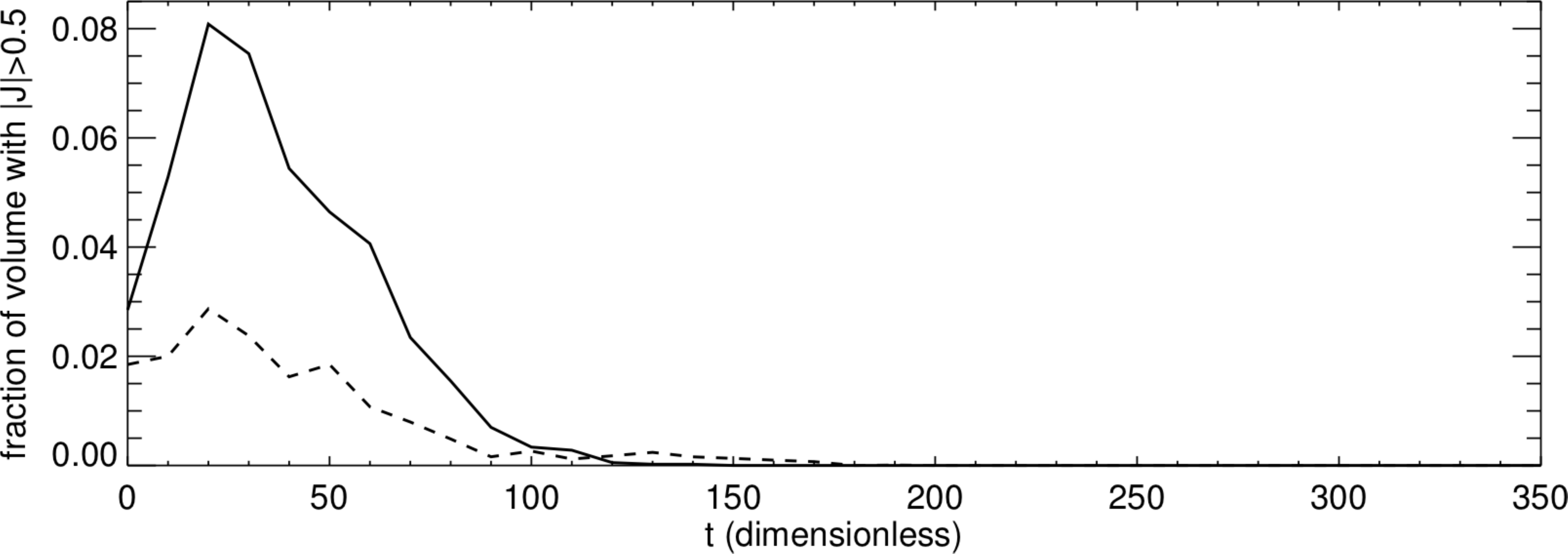}

\caption{Upper panel:   Maximum value of the current density $\vert {\bf J} \vert$ in the domain with time.
 Lower panel: Fraction of the volume over which $\vert {\bf J} \vert > 0.5$.
In both cases the solid line corresponds to $E^{3}$ and the dashed line to $S^{3}$.}
\label{fig:jmaxes}
\end{figure}

In qualitative terms the resistive MHD evolutions of $E^{3}$ and $S^{3}$ are somewhat similar.
The large-scale currents of the initial state (Figure~\ref{fig:ICrmhd}) collapse towards
two thin current layers for $E^{3}$ and one thin current layer for $S^{3}$.
This process is shown in Figures~\ref{fig:JisoS}, \ref{fig:JcontoursE} and \ref{fig:JcontoursS}.
It is at this early stage that the  strongest currents  are present in the system (see Figure~\ref{fig:jmaxes}, upper panel).
As time progresses more current layers are found which are, individually, weaker in their intensity. 
For $E^{3}$ a large number of current layers appear and they  have a volume-filling effect (see, for example, $t=50$
in the relevant figures).  For $S^{3}$, by contrast, a smaller number of current layers are present and these 
are mainly located in the central regions of the domain (with respect to $x$ and $y$).
 To quantify this difference we show in Figure~\ref{fig:jmaxes} (lower panel) the fraction of the volume
(in $[-4,4]^{2} \times [-24,24]$)  for which $\vert {\bf J} \vert > 0.5$ for both $E^{3}$ (solid line) and $S^{3}$ 
(dashed line).  The quantity is consistently larger for $E^{3}$, reaching a maximum of $8.1\%$, compared 
with only $2.8\%$ for $S^{3}$.
In the later stages of the evolution the systems move towards equilibria containing only large-scale, weak currents.  
In the case of $E^{3}$ these form two vertical tubes of current while for $S^{3}$ only one current tube is present.
Note that we have run these simulations for a long time (up to $t=650$) and find the state reached at $t=350$ 
represents the end-state since after this time only slow changes in the magnetic field take place and these
are due to global diffusion.

\begin{figure}[tbp] 
\centering
\includegraphics[width=0.3\textwidth]{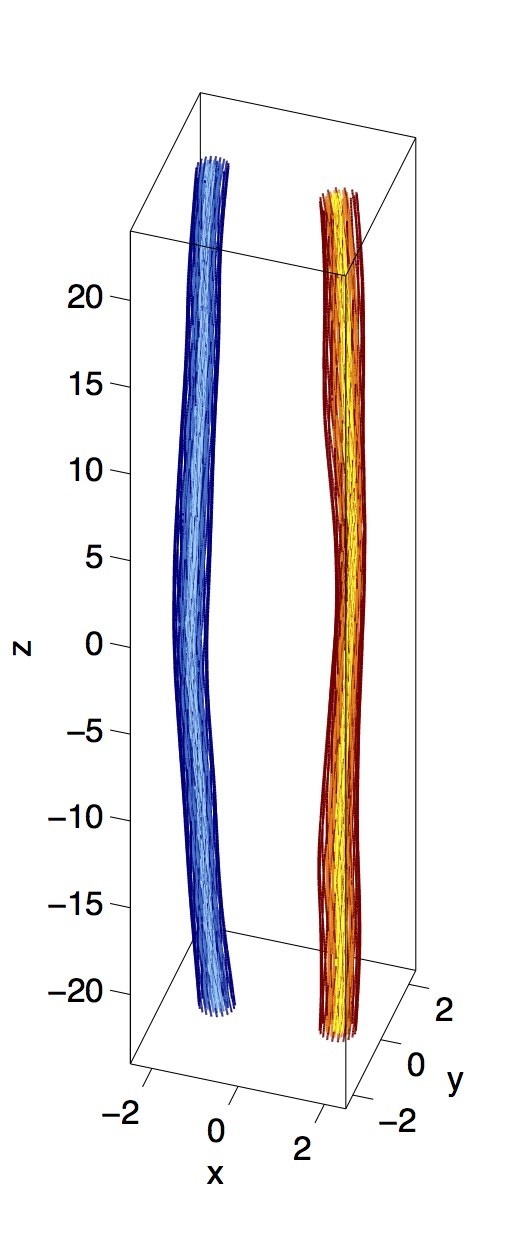}
\includegraphics[width=0.3\textwidth]{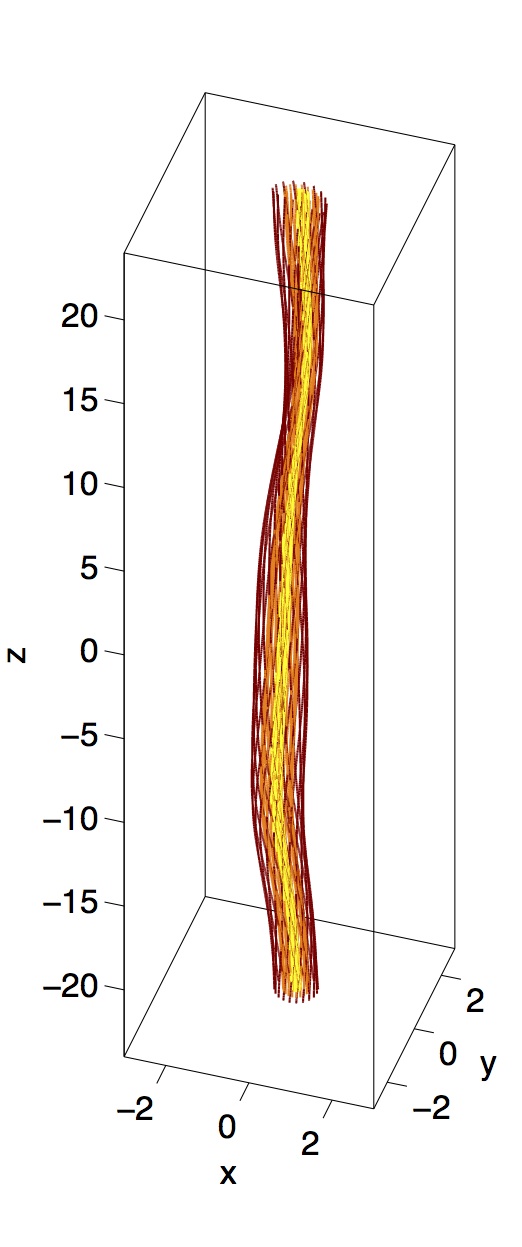}
\caption{Some illustrative field lines in the final states for $E^{3}$ (left) and $S^{3}$ (right).
The field lines have been chosen to show the fundamental structure of the states, two
 flux tubes of opposite twist for $E^{3}$ and a single twisted flux tube for $S^{3}$.}
\label{fig:finalstate}
\end{figure}

\begin{figure}[tbp] 
\centering
\includegraphics[width=0.28\textwidth]{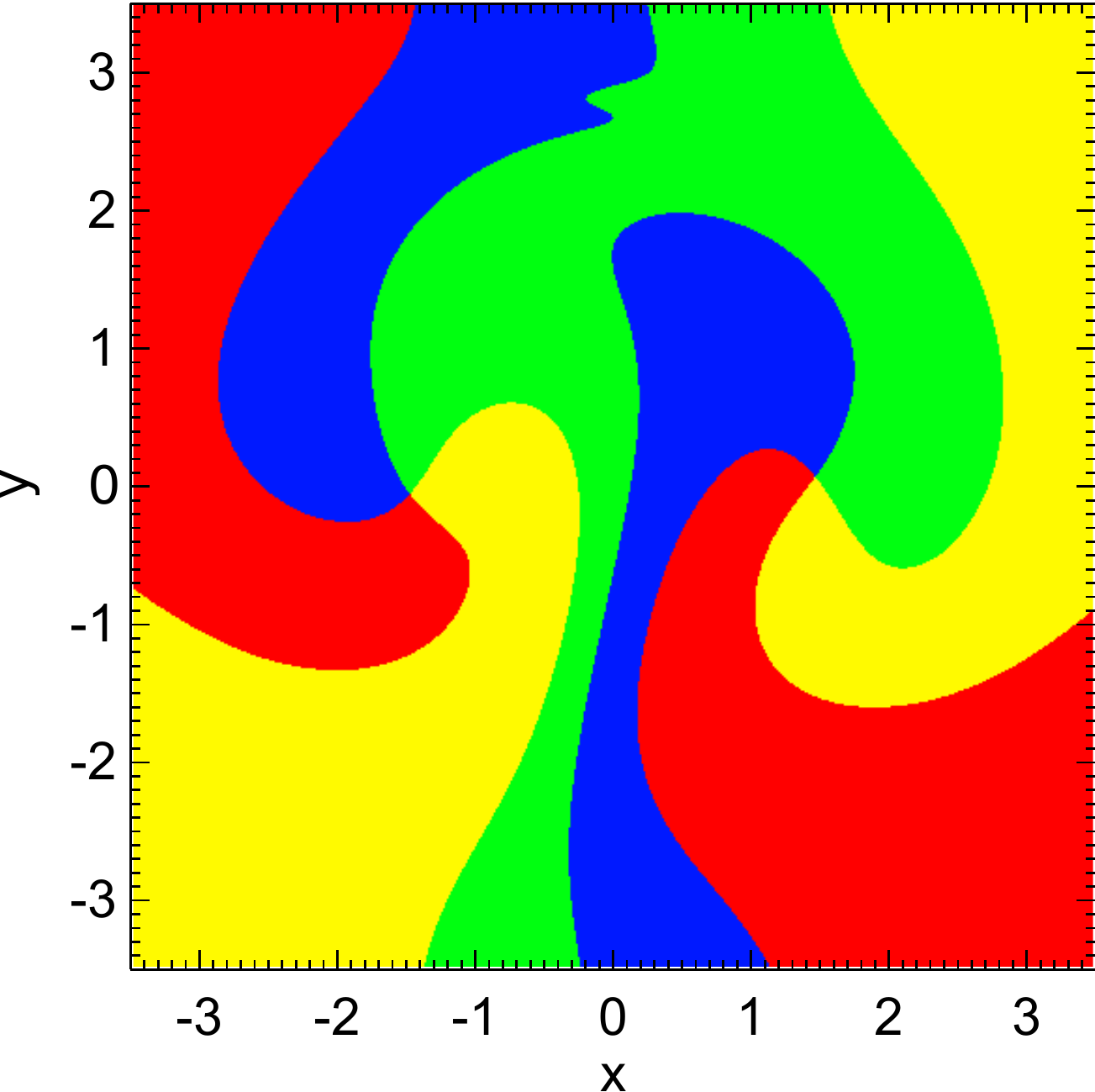} 
\includegraphics[width=0.28\textwidth]{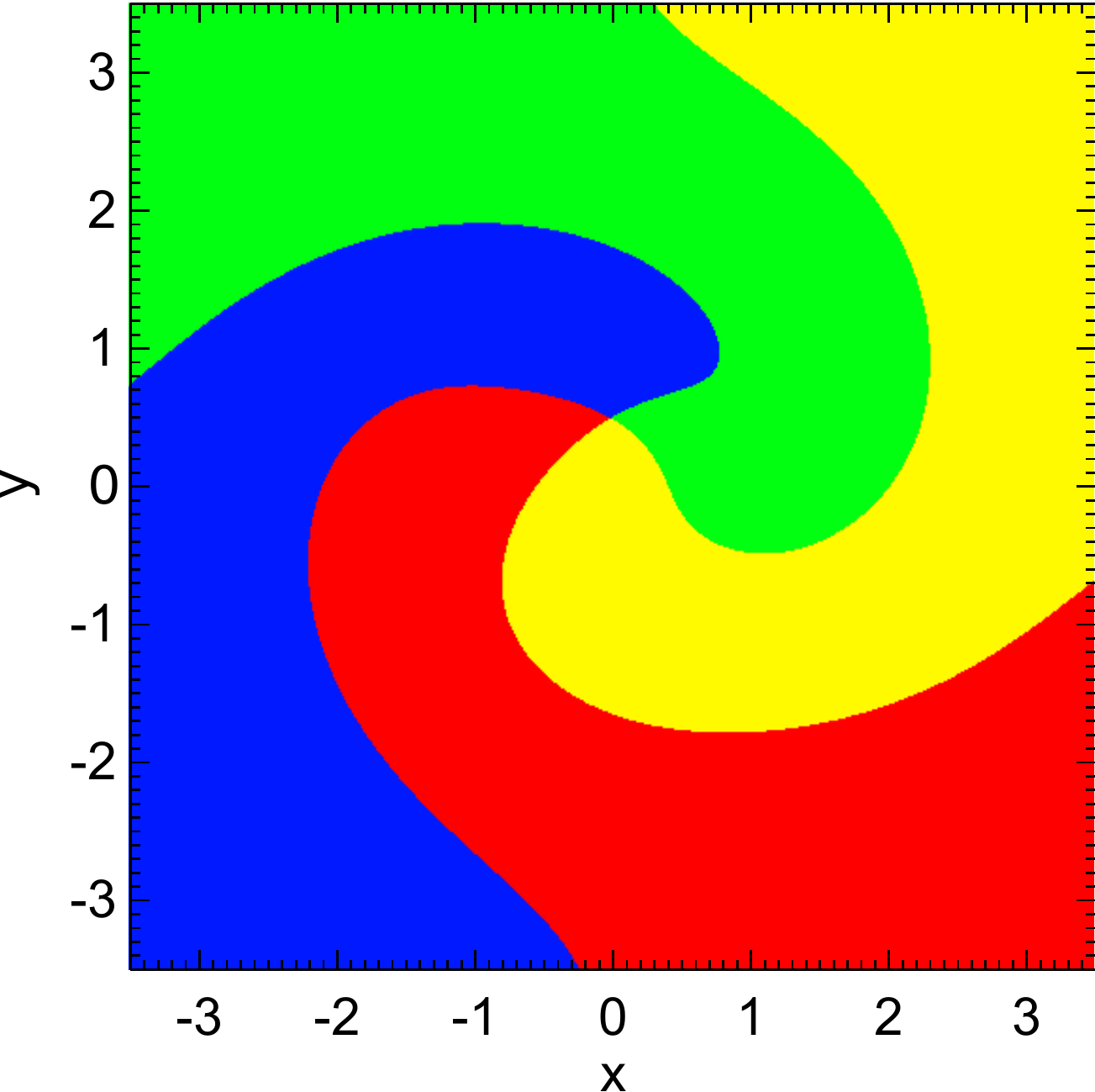} 
\caption{Colour maps in the final states of the resistive MHD simulations
for $E^{3}$ (left) and $S^{3}$ (right).}
\label{fig:colormaps2}
\end{figure}

Considering the final states of the resistive relaxations, although the number of current tubes present in each case 
is the same as in the  respective initial states, the  structure of the magnetic fields is very different.
The end state for $E^{3}$ consists of two unlinked magnetic flux tubes of opposite sign of twist, as shown in
the left-hand image of Figure~\ref{fig:finalstate}.  The end state for $S^{3}$, by contrast, consists of a single 
magnetic flux tube of positive twist as shown in the right-hand image of Figure~\ref{fig:finalstate}.
The particular field lines plotted have been chosen because they portray the
nature of the corresponding continuous magnetic fields.  We show this nature in two ways, firstly in Figure \ref{fig:colormaps2}
using the colour-map technique described in Section~\ref{sec:compare}.
The corresponding colour maps for the initial states of both fields are shown in Figure \ref{fig:colormaps}.
The first, striking, feature seen is the simplicity of the colour-maps in the final states in comparison to 
the initial states.  This simplification comes from the un-braiding of the coronal loops to form two flux tubes for
$E^{3}$ and one for $S^{3}$.   The difference in number of flux tubes may also be seen in the colour maps, there
being two distinct intersections of all four colours (periodic orbits) for $E^{3}$ and only one for $S^{3}$.  These
intersections mark the centre of the flux tube axes.

A second feature of the colour map tells us why we cannot relax to a single flux tube in the $E^{3}$ simulation
(as determined by Yeates {\it et al.}~2010 {, Yeates \& Hornig 2011}).  If we walk around the boundary of the domain for $E^{3}$ then
we meet each of the four colours twice (and in the order red--yellow--green--blue).  
This is because the overall topological degree of the state is $+2$.  This is the same boundary pattern that
was present in the initial state of $E^{3}$ since the relaxation only affects the interior of the domain; 
mathematically, the topological degree of the field is conserved in the relaxation.  Furthermore, a topological degree of $+2$ is inconsistent with a single flux tube. 
Considering $S^{3}$, the boundary of the domain shows each colour appearing only once (but in the same sequence,
topological degree $+1$) and so relaxation to a single flux tube is possible.

\begin{figure}[tbp] 
\centering
\includegraphics[width=0.33\textwidth]{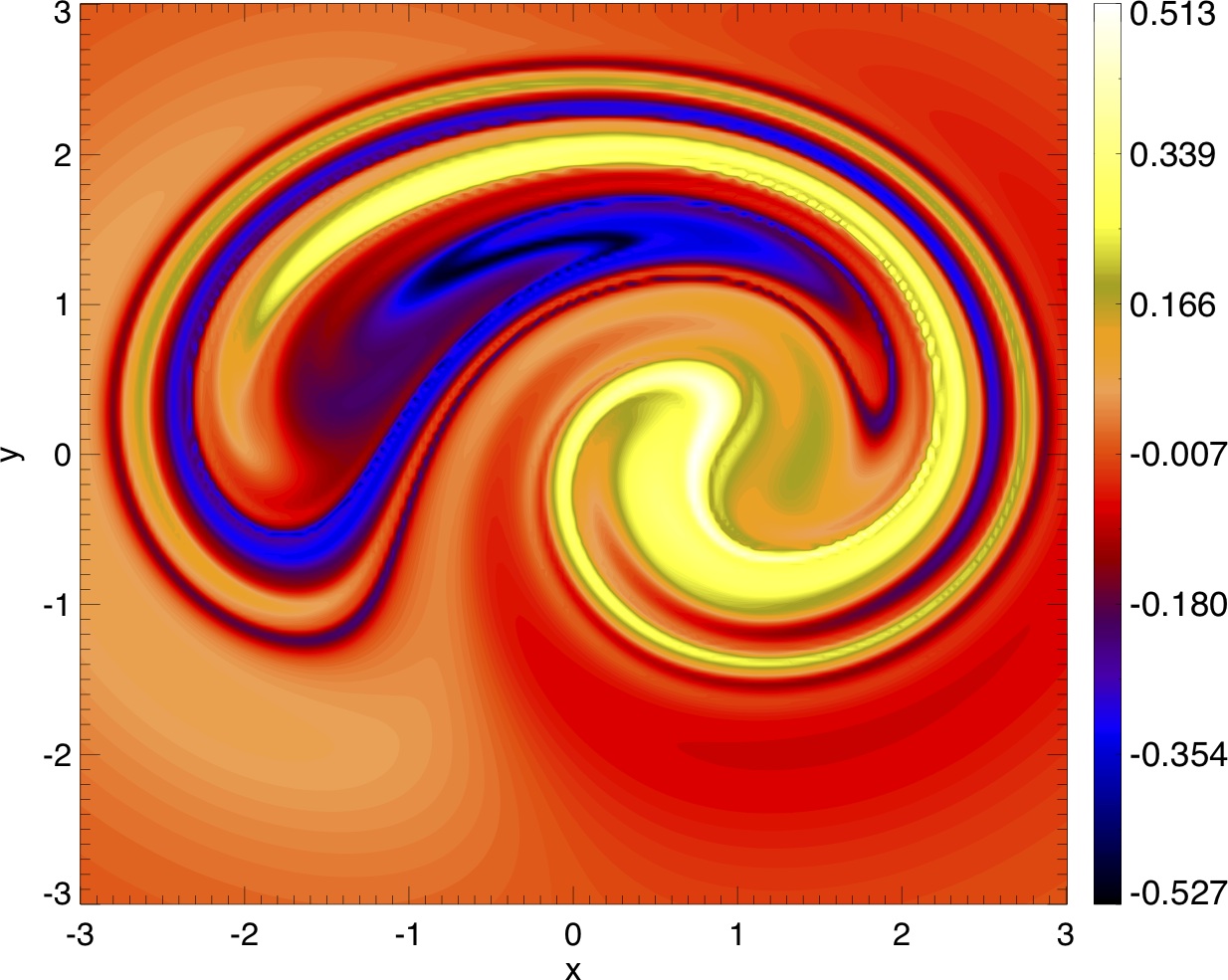}
\includegraphics[width=0.33\textwidth]{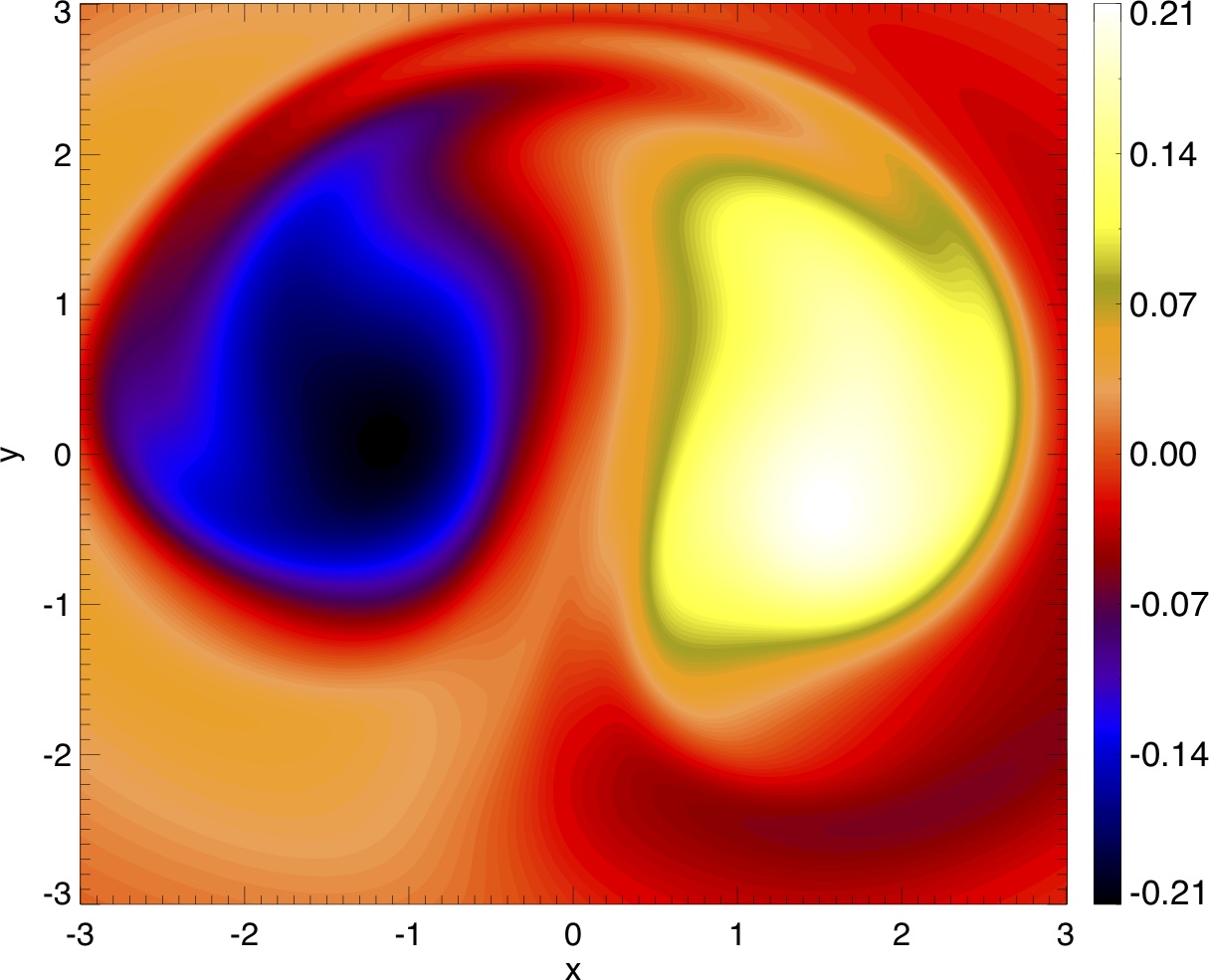} \\
\includegraphics[width=0.33\textwidth]{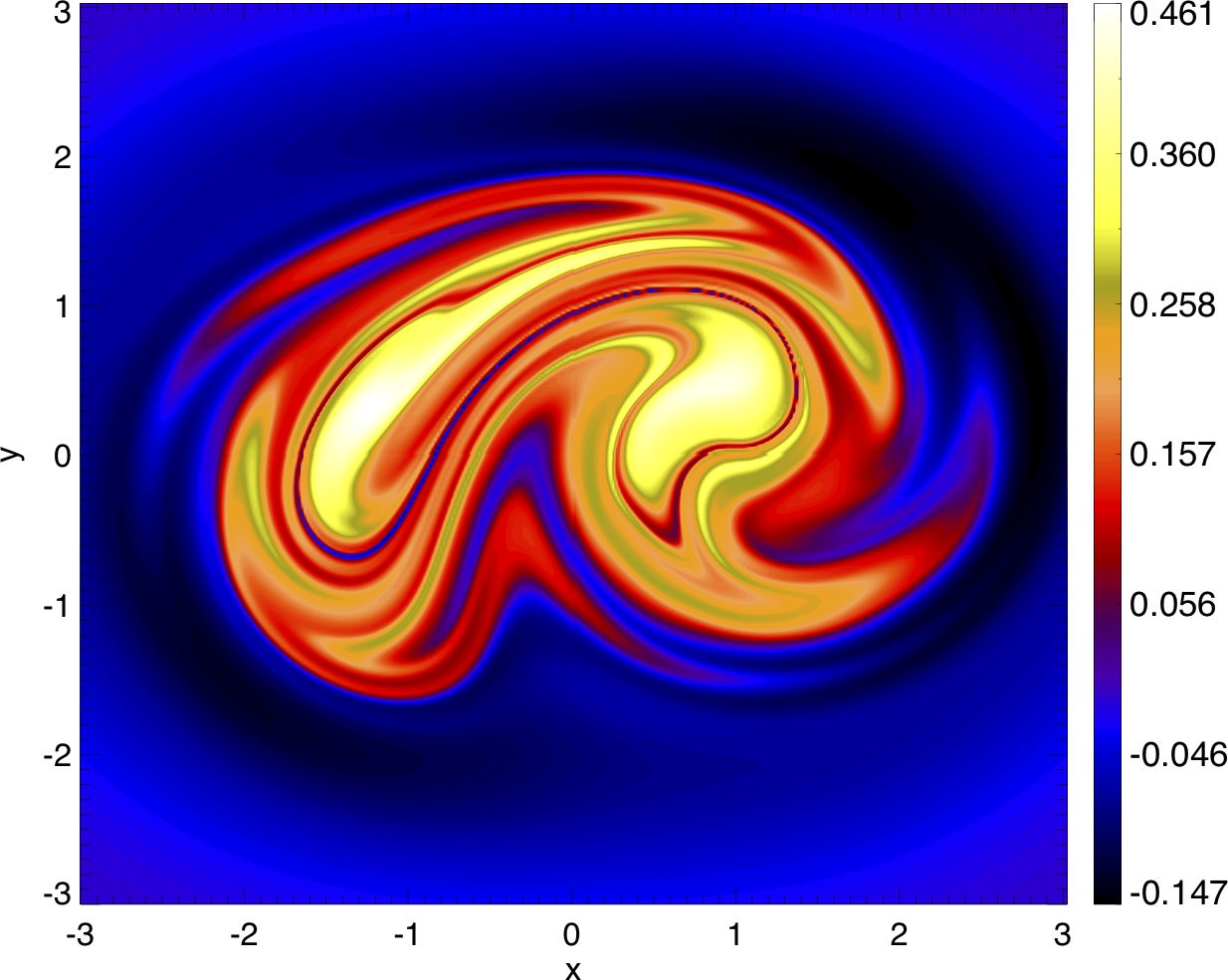}
\includegraphics[width=0.33\textwidth]{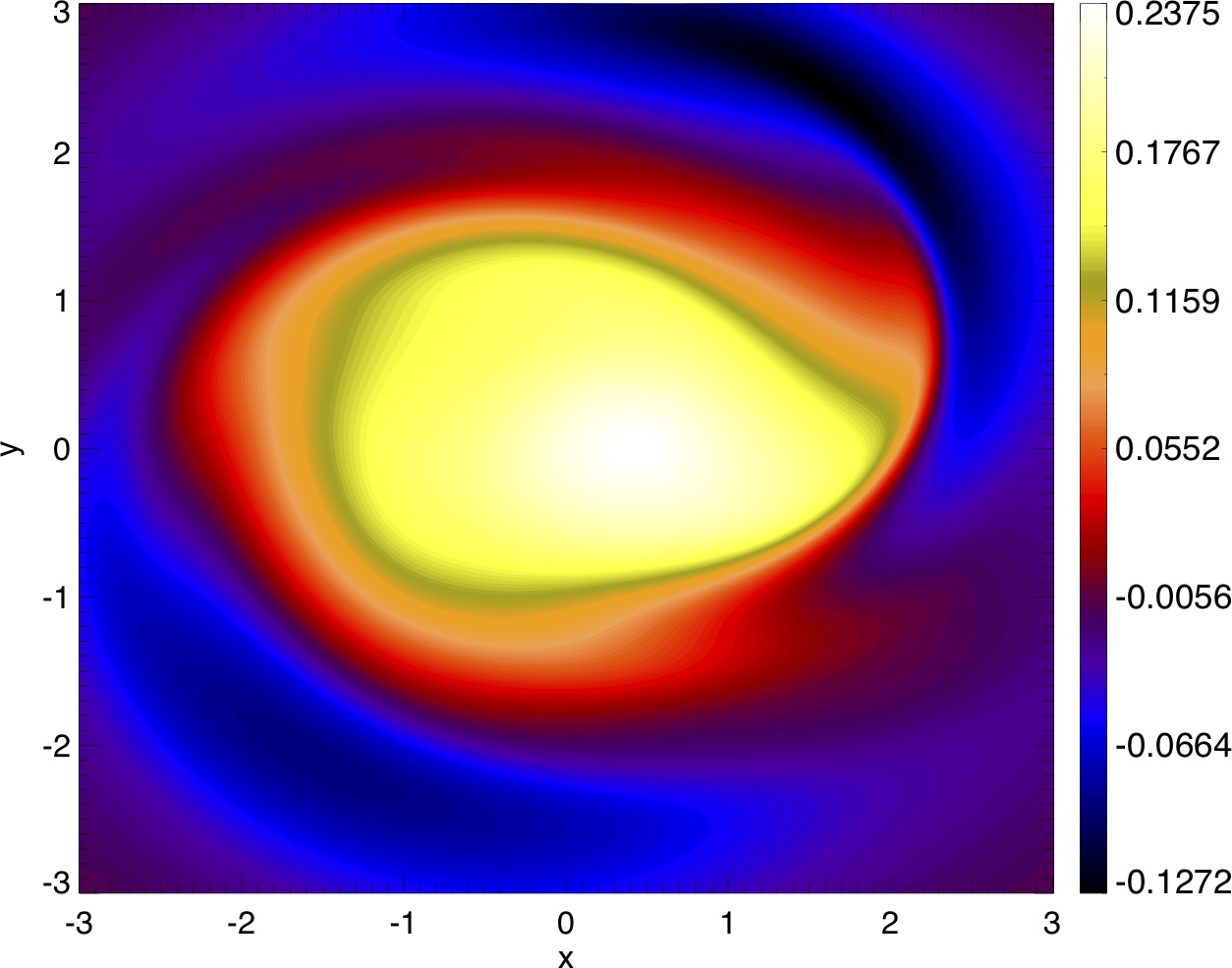}
\caption{Contours of the quantity $\bar{\alpha^{*}}$
averaged along field lines in the initial (left) and final (right) states for $E^{3}$ (upper panel) 
and $S^{3}$ (lower panel).}
\label{fig:meanalpha}
\end{figure}

A second way to examine the nature of the final state is to look at the mean value along field lines,
$\bar{\alpha^{*}}$, of the quantity $ {\alpha^{*}} = {{\bf J}\cdot {\bf B}}/{{\bf B}\cdot {\bf B}}$.  
Figure~\ref{fig:meanalpha} shows $\bar{\alpha^{*}}$ on  the lower boundary of the domain for both the initial and 
final states of the resistive MHD simulations.
For a perfectly force-free field, $\bar{\alpha^{*}}$  {is simply} the force-free parameter $\alpha$.
In our simulations the initial state is only approximately force-free and the finite gas pressure gives relaxed 
fields that are also not perfectly force-free.  In both cases $\bar{\alpha^{*}}$ is the appropriate quantity to consider. 
The images in the initial state further illustrate the complexity of the magnetic field
while the images for the final state confirm them as being much simpler. 
The basic un-braiding into two ($E^{3}$) or one ($S^{3}$) flux tubes is clearly shown in  $\bar{\alpha^{*}}$.

Neither final state could be considered as a globally linear-force--free field 
but the individual twisted flux tubes do lie in regions of approximately constant  $\bar{\alpha^{*}}$. 
 {Thus while $E^3$ is clearly in contradiction to a `Taylor-like' relaxation (the Taylor
state in that case being the uniform field), one could argue that the final state  for $S^3$ has similarities with a 
Taylor state if we restrict ourselves to a domain over which significant current fragmentation occurs. 
For further considerations on the relevant Taylor state for $S^3$ we must consider the value of the total 
helicity for the field since this quantity is invoked as the constraint determining that state.}

 {
Total helicity is used as a constraint under Taylor relaxation theory since, although it is}
 not an invariant in a resistive MHD evolution, it is thought to be approximately conserved on the relaxation 
 timescales.   {To confirm this for these simulations we calculate the total relative helicity in the
 initial and final states for both fields. Here we use the reference field  $1 \mathbf{e}_{z}$, 
 which is the potential field satisfying the same boundary conditions. 
  (Note that this is equivalent 
 to calculating the total helicity in the torus obtained by identifying the top and bottom boundaries of our domain  and assuming no flux links the hole of the torus.) 
In the initial states we find $H(E^{3}, t=0) = 0.0$ and $H(S^{3}, t=0) = 349.6$.
while for the final states we find  {$H(E^{3}, t=350) = 1.2$} and $H(S^{3}, t=350) = 350.9$.
Thus the conservation in both cases is excellent with  {only a small production of helicity 
(by magnetic reconnection) in each case.}}

 {To see whether the final state for $S^{3}$ can be considered a Taylor state on the domain
on which the current fragmentation occurs (see Figure~\ref{fig:JcontoursS}) we calculate  {the Taylor} state on a cylindrical domain with the relevant parameters 
(radius $r=3$, vertical flux of $9\pi$ and  total helicity $350.8$).  
The result is an axially symmetric Lundquist solution (Lundquist 1951)
with constant $\alpha=0.114$ so providing a reasonable qualitative and quantitative match to our results.}

Having described the basics of the resistive relaxations we proceed to consider the energetics of the process.
This is of interest for determining whether and how relaxation processes such as these may heat the solar
corona.

\subsection{Energy and heating}
\label{sec:energyheating}

\begin{figure}[htbp] 
\centering
\includegraphics[width=0.55\textwidth]{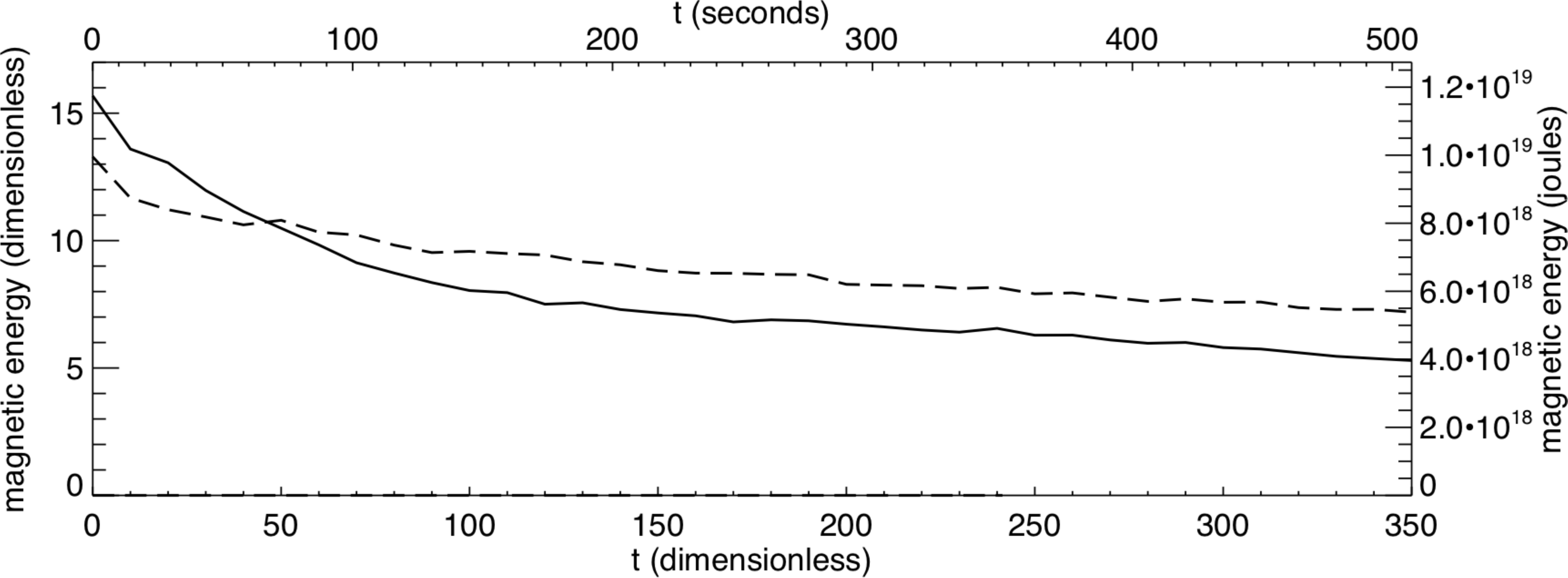}

\vspace*{0.5cm}

\includegraphics[width=0.55\textwidth]{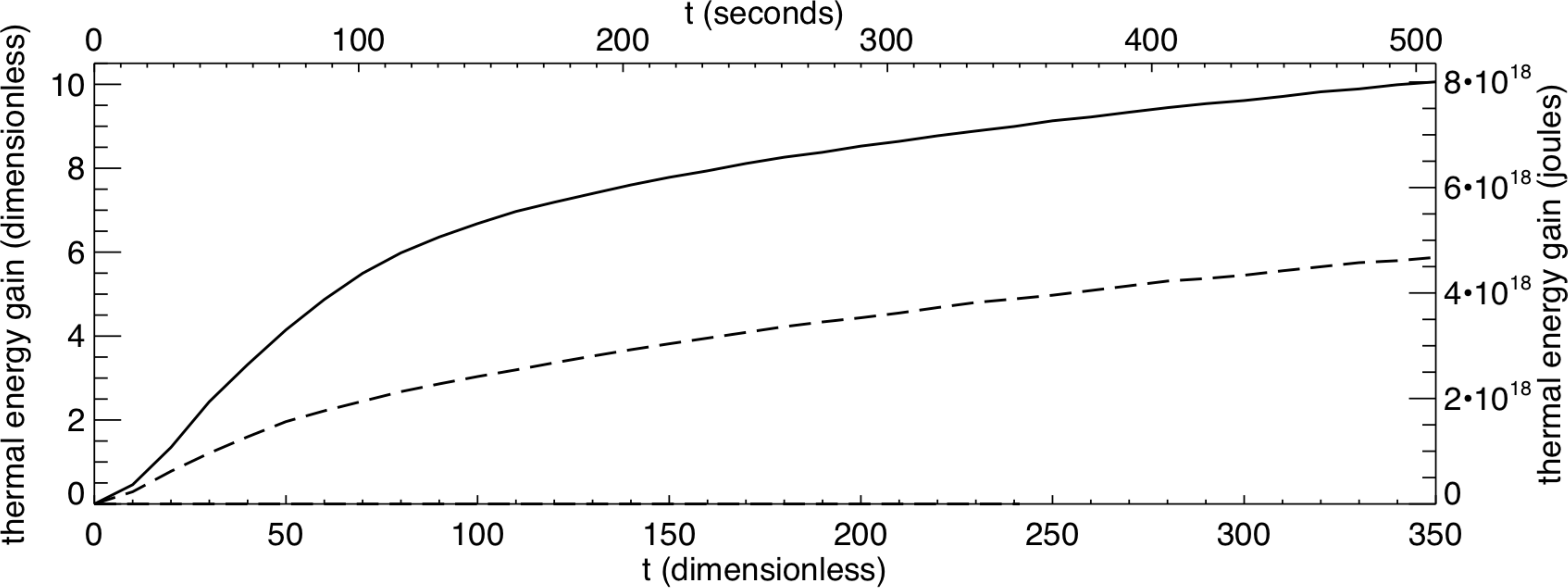}
\caption{ {The total magnetic energy in excess of potential (upper image) 
and the total thermal energy above that of the initial state  (lower image).
In both cases the solid line is for $E^{3}$ while the dashed line is for $S^{3}$.}}
\label{fig:energies}
\end{figure}

In order to provide a solar coronal interpretation for these two relaxation events we present our 
results in dimensional terms.  To do so we choose characteristic values for the magnetic field strength as
$B_{0} = 10\textrm{ G}$, the unit of length as $l_{0} = 1 \textrm{ Mm}$ and the electron number density as 
$10^{15} \textrm{ m}^{-3}$.  Accordingly we have a unit of time as $t_{0} = 1.45 \textrm{ s}$,
temperature as $T_{0} = 3.44 \times 10^{7} \textrm{ K}$ and velocity as $v_{0} = 689.7 \textrm{  km s}^{-1}$.
This gives our initial magnetic loops for both $E^{3}$ and $S^{3}$  a temperature 
of $2.30 \times  10^{6} \textrm{ K}$ and a loop length of $48 \textrm{ Mm}$.
Although the overall time of the relaxation then corresponds to $507$ seconds,
in considering this timescale it is important to note that the time taken for the full relaxation
has previously (Pontin {\it et al.}~2011) been shown to be dependent on the value of the resistivity, 
with relaxation time increasing with decreasing resistivity.   The resistivity taken as within reach of
computing resources available to us remains several orders of magnitude too high when compared to
 assumed coronal values. 
 Accordingly, the timescale for a real relaxation is likely to be greater than $500$ s.

\begin{figure*}[tbp] 
\centering
\includegraphics[width=0.9\textwidth]{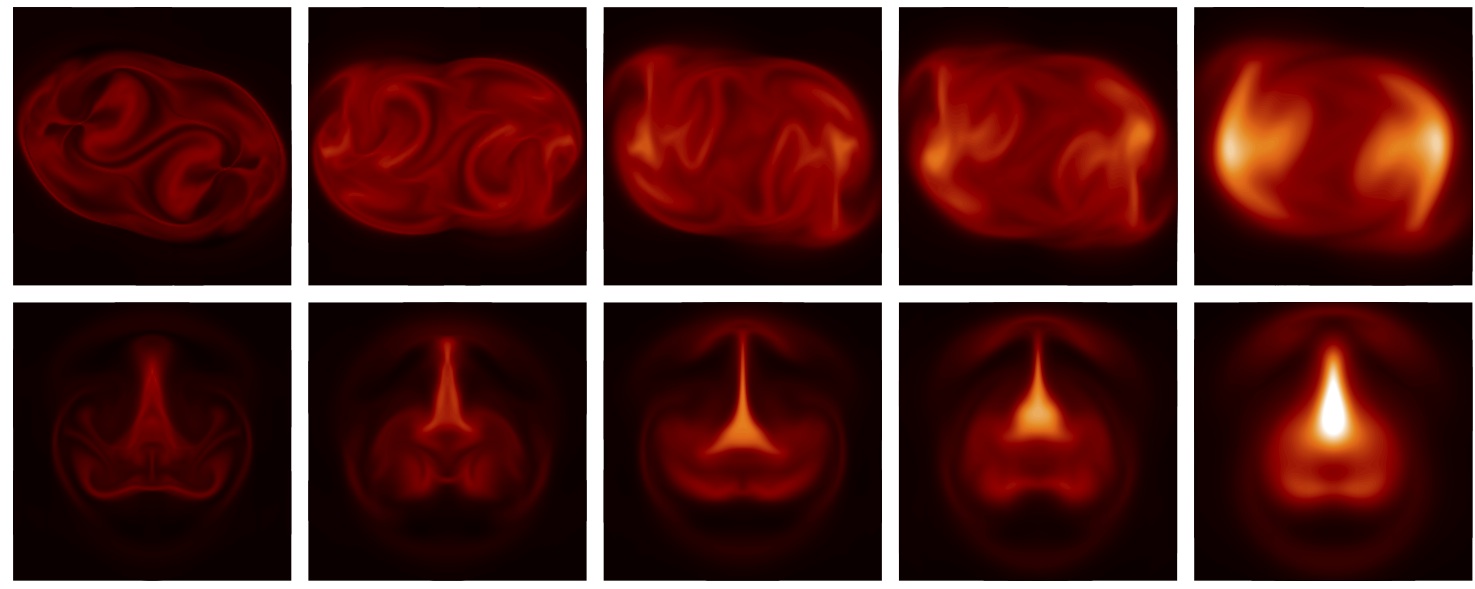}\\ [10pt]
\includegraphics[width=0.4\textwidth]{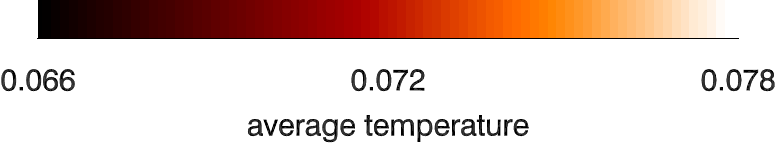}~~~~~~
\includegraphics[width=0.4\textwidth]{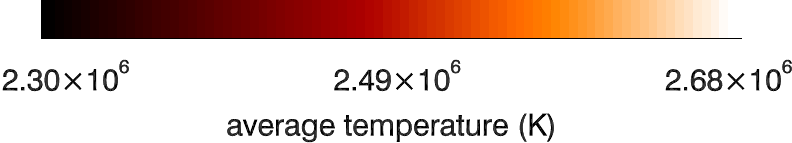}
\caption{Average temperature along field lines at their intersection with the $z=0$ plane
($x,y, \in [-3,3]$) for $E^{3}$ (upper panel) and $S^{3}$ (lower panel) at 
dimensionless times $t=40 $, $80$, $120$, $180$ and $350$.}
\label{fig:temperatures}
\end{figure*}

 {The first thing that we note is that during the initial ideal relaxation process, the field $S^{3}$ is able to 
decrease its magnetic energy by more than $E^{3}$. Specifically, both $E^{3}$ and $S^{3}$ 
 begin with $3.76 \times 10^{19} J$ of excess energy prior to the ideal relaxation but, after this 
ideal phase, $E^{3}$ has $1.25\times10^{19}J$ remaining in excess of the potential field while 
$S^{3}$ has  $1.06\times10^{19}J$.   This is a result of the higher topological 
complexity of $E^{3}$ -- the highly braided field line mapping places a greater constraint on the ideal relaxation. 
Only once the resistive relaxation is begun, and the field begins to simplify, can this energy be liberated.}

 {The decay of magnetic energy in excess of potential
during the resistive relaxation is shown in Figure 12 (upper image).}
 {While in the initial states for these resistive simulations $E^{3}$ contained marginally more magnetic energy,
over the course of the relaxation it releases $8.27 \times 10^{18} J$ of magnetic energy
while $S^{3}$ releases only $4.85 \times 10^{18} J$.}  In terms of the fraction of 
 {free magnetic energy present at the start of each of the resistive simulations,}
 $66.2\%$ is released for  $E^{3}$ but only $45.88\%$ for $S^{3}$.
That is, $E^{3}$ has been much more efficient at releasing energy.  We claim this is because the higher degree
of complexity in its initial state (as measured by the topological entropy) results in a much more fragmented,
volume filling, system of current sheets (Figures~\ref{fig:JisoS}, \ref{fig:JcontoursE} and \ref{fig:JcontoursS})
and so for a more efficient relaxation.

Much of the magnetic energy released is converted into thermal form.  The evolution of this quantity is shown in 
Figure~\ref{fig:energies} (lower image); we find an increase in thermal energy of $8.01 \times 10^{18} J$ for 
$E^{3}$ and  $4.68 \times 10^{18} J$ for $S^{3}$.  The remaining energy takes the form of some residual 
non-zero kinetic energy in the final states of both systems.  This is a result of large-scale weak oscillations 
in the flux tubes.  

We now examine changes in the temperature of both the $E^{3}$ and $S^{3}$ loop systems during the relaxations.
In doing so we must bear in mind that the initial plasma beta, $\beta (t=0) = 0.133$, is relatively high
and so temperature increases will be lower than if a  lower plasma beta were used.  Another factor is the 
lack of cooling terms in the energy equation which gives a competing effect in maintaining 
an artificially high temperature.  Nevertheless we consider it useful to examine features of the temperature evolution
in order to determine general trends.

The initial temperature in both cases is $2.30 \times 10^{6}K$.  In $E^{3}$ the peak temperature increases
to $2.76 \times 10^{6} K$ and in $S^{3}$ to $2.92 \times 10^{6} K$.  These are increases of factors $1.2$ and $1.27$ 
respectively. 
We have re-run the $E^{3}$ simulation for an initial plasma beta an order of magnitude lower ($\beta (t=0)=0.0133$)
and found an increase  {in maximum}  temperature of a factor $3.1$ in that case.
 While the lack of realistic energy transport or loop stratification 
prevents us from  making  {more} realistic statements
about heating along the loops, {we can analyse the  integrated temperature along field lines. This is justified by the excellent heat conduction along the magnetic field lines in the corona  compared to the almost zero heat conduction perpendicular to the field.
 The quantity provides insight into the space-filling nature of the heating.} {The field line average temperature} is shown in Figure~\ref{fig:temperatures} in the $z=0$ plane and for various stages of the evolution 
where the colour-scale for all images is normalised to the  {same} maximum value 
(the end-state of $S^{3}$).
Some clear findings emerge here.  The spatial distribution of high temperature is much more homogenous for 
$E^{3}$.  Nevertheless two brighter features are seen in the end-stages, corresponding to the two separated flux 
tubes in the end-state.  In the early evolution several `hot-spots' are present in the loop and these change location 
in time.  We assert that the heating would be even more homogenous for lower resistivity as more current sheet
fragmentation would be achieved (as documented in Pontin {\it et al.}~2011).
For $S^{3}$ heating primarily takes place in a thin layer in the centre of the domain with the temperature here being
slightly higher   {than anywhere in the domain for $E^{3}$}
since the heating is less distributed.

These contrasting behaviours in temperature distribution arise from a basic property of the initial 
configurations, namely, the  {difference in} topological entropy.   
With  higher topological entropy, $E^{3}$ evolves through a highly fragmented system
of current layers which allow for a uniform loop heating.  The field $S^{3}$, with lower topological entropy,
evolves with fewer current layers which are more patchy in their distribution.  The temperature profile 
is then correspondingly less homogeneous.

\section{Conclusions}
\label{ref:conc}

In this paper we have shown that the ability of non-ideal processes to heat coronal loops via magnetic
reconnection is crucially dependent on the nature of photospheric motions at the footpoint of the loops.
Even for photospheric motions injecting the same amount of magnetic energy into loop systems, 
the magnetic field configurations generated can lead to quite different amounts of energy being released
in a relaxation event.

Two effects influence the amount and the distribution of heating. 
Firstly the heating is affected by the amount of energy that can be released in a relaxation 
on dynamic time-scales.  Relaxation preserves the total helicity (Taylor, 1974) and possibly also further 
topological constraints (Yeates {\it et al.}~2010).  Here, large amounts of magnetic helicity prove to be 
counter--productive as they raise the energy of the allowable end-state.  In general more complex, 
non-coherent, braiding of the field lines allows for greater subsequent energy release.
Secondly, the ability to release stored magnetic energy relies on a sufficiently fragmented, 
volume--filling system of current layers.  Fragmentation is more efficient  with more complex
 braiding, and also increases the homogeneity of the heating.

f Note that the braid complexity (measured here by the topological entropy) is not entirely
independent of the total helicity.  High values of total helicity over a domain require 
a coherence in the structure that limits the topological entropy.  Conversely, any non-trivial magnetic 
field on a closed or periodic domain must have a non-vanishing helicity density and so one can always
find a sub-domain of such a  field over which the total helicity is non-zero.  Formalising a relationship
between the two effects is beyond the scope of the present study.

In order to support  our  claims we have presented simulations of the resistive MHD evolution of two magnetic loops.
Both loops can be generated by pairs of rotational photospheric motions acting on footpoints of a uniform field
and contain the same amount of magnetic energy.
For our first field, $E^{3}$, the two basic motions are in opposite directions while for our second field, $S^{3}$, both 
motions are in the same direction.   The boundary motions for $E^{3}$ lead to a field that is highly braided.
The complexity in the field line mapping is visually apparent in a map of the squashing factor, $Q$ 
(Figure~\ref{fig:Qinitial}) and can be quantified by the topological entropy.
The boundary motions for $S^{3}$ lead to a simpler pattern of field line connectivity with 
a simpler $Q$-structure (but similar maximum $Q$) and a lower topological entropy.

The fields both undergo resistive relaxations on a fast, Alfv{\' e}nic timescale.  The relaxation for $E^{3}$ 
is more efficient;  more current  sheets are generated which have a greater volume filling tendency. By the 
end of the relaxation the {free} magnetic energy of the field has been reduced by $\sim 65\%$ 
with this being converted primarily to thermal energy.  The entire loop is heated in a relatively homogenous manner.  
The efficiency of the relaxation arises from the high topological entropy of the initial state.
The relaxation for $S^{3}$, with its lower topological entropy, is less efficient.    
A smaller number of current sheets result and these have a lower volume filling tendency.  The free magnetic 
energy of the field is reduced by only $\sim 45\%$.  The heating of the loop takes place less uniformly, in one 
central location.

 {Extrapolating to the solar corona, this mechanism provides a way to deposit heat in a spatially uniform 
way throughout the entire body of a coronal loop since complex braiding patterns of the field lead to a 
relaxation through a complex, space-filling set of current sheets.
In order to determine how important this mechanism might be in practice, it would be useful to look at very 
high resolution data of surface motions (by way of fragment tracking) to consider the 
topological entropy of photospheric motions in different regions of the Sun over the solar cycle.
Furthermore, a number of extensions to the models presented here should be made
to allow for more detailed predictions relating to energetics.  These include a realistic stratification
of the model atmosphere and additional physics to the energy equation.}

\vspace*{0.25cm}

\noindent
{\small {\it Acknowledgements.}
The authors would like to thank an anonymous referee whose comments helped to 
improve this paper.
Simulations were carried out on the SRFC and SFC (SRIF) funded linux clusters of the UKMHD consortium.
G.H. and A.R.Y. acknowledge financial support from the UK's STFC (grant number ST/G002436).}

\newpage

\noindent
{\bf References}

\noindent
Aschwanden, M.J., ApJ {\bf 634}, L193 (2005).

\noindent
Aschwanden, M.J., Boerner, P.,   ApJ {\bf 732}, 81 (2011).

\noindent
Bingert, S., Peter, H., A\&A {\bf 530}, A112 (2011).

\noindent
Birman, J.S., `Braids, Links and Mapping Class Groups', Annals of Mathematics 
Studies, \newline \hspace*{0.2cm}  Princeton University Press (Princeton, New Jersey) (1974).

\noindent
Craig, I.J.D., Sneyd, A.D.,  ApJ {\bf 311}, 451 (1986).

\noindent
Craig, I.J.D., Sneyd, A.D., Solar Phys. {\bf 232} 1-2, 41 (2005).

\noindent
De Moortel, I., Galsgaard, K., 
A\&A {\bf 451} 1101 (2006).

\noindent
De Moortel, I., Galsgaard, K.,  A\&A {\bf 459} 627 (2006).

\noindent
Dixon, A.M., Berger, M.A., Browning, P.K., Priest, E.R.,  A\&A {\bf 225}, 156 (1989).

\noindent
Galsgaard, K., Nordlund, A.,  JGR {\bf 101} A6, 13445 (1996).

\noindent
Gold, T., in The Physics of Solar Flares, ed. w. Hess (NASA SP-50) p.389 (1964).

\noindent
Gudiksen, B.V., Nordlund, {\AA}, ApJL 1, L113 (2002).

\noindent
Heyvaerts, J., Priest, E.R.,  A\&A {\bf 137}, 63 (1984).

\noindent
Hood, A.W., Browning, P.K., Van der Linden, R.A.M.,  A\&A {\bf 506}, 913 (2009).

\noindent
Klimchuck, J.A.,  Solar Phys. {\bf 193}, 53 (2000).

\noindent
Klimchuk, J.A.,  Solar Phys., {\bf 234}, 41 (2006).

\noindent
Kusano, K.,  ApJ {\bf 631}, 1260 (2005).

\noindent
Longbottom, A.W., Rickard, G.J., Craig, I.J.D., Sneyd, A.D.,  ApJ {\bf 500} 471 (1998).

\noindent
Longcope, D.W., Sudan, R.N., ApJ {\bf 437}, 491 (1994).

\noindent
L{\'o}pez Fuentes, M.C., Klimchuk, J.A.,  D{\'e}moulin, P.,  ApJ {\bf 639}, 458 (2006).

\noindent
L{\'o}pez Fuentes, M.C., D{\'e}moulin, P., Klimchuk, J.A.,  ApJ {\bf 673} 1, 586 (2008).

\noindent
Lundquist, S.,  Arkiv Fysik {\bf 2}, 361 (1951).

\noindent
 {Moussafir, J.-O.,  Func. Anal. Other Math. {\bf 1}, 37 (2006).}

\noindent
Nandy, D., Hahn, M., Canfield, R.C., Longcope, D.W.,  ApJ {\bf 597}, L76 (2003).

\noindent
 {Newhouse, S., Pignataro T., J. Stat. Phys. {\bf 72}, 1331 (1993).}

\noindent
Ng, C.S., Bhattacharjee, A., Phys. Plasmas {\bf 5} 11, 4028 (1998).

\noindent
Patsourakos, S., Klimchuk, J.A., ApJ {\bf 689}, 1406 (2008).

\noindent
Parker, E.N., Cosmical Magnetic Fields: their origin and their activity.   \newline
\hspace*{0.2cm} Oxford: Clarendon Press (1979).

\noindent
Parker, E.N., Spontaneous Current Sheets in Magnetic Fields. \newline 
\hspace*{0.2cm}  Oxford: Oxford University Press (1994). 

\noindent
Peter, H., Gudiksen, B.V., Nordlund, \AA., ApJL {\bf 617}, L85 (2004).

\noindent
Polymilis, C., Servizi, G. Skokos, Ch., Turchetti, G., Vrahatis, M.M.,
Chaos {\bf 13} 94 (2003).

\noindent
Pontin, D.I., Hornig, G., Wilmot-Smith, A.L., Craig, I.J.D., ApJ, 700, pp. 1449 (2009).

\noindent
Pontin, D.I., Wilmot-Smith, A.L., Hornig, G., Galsgaard, K.,  A\&A 525 A57 (2010).

\noindent
Reale, F.,  Living Rev. Solar Phys. {\bf 7}, 5 (2010).  \newline \hspace*{0.2cm} 
[Online Article]: cited 16/08/2011, http://www.livingreviews.org/lrsp-2010-5

\noindent
{Schmelz}, J.~T., {Nasraoui}, K., {Rightmire}, L.~A., {Kimble}, J.~A., {del Zanna}, G.,
{Cirtain}, J.~W., {DeLuca}, E.~E., {Mason}, H.~E., 
 ApJ {\bf 691}, 503 (2009).

\noindent
Taylor, J.B.,  Phys. Rev. Lett. {\bf 33} 1139 (1974).

\noindent
Taylor, J.B., Rev. Mod. Phys. {\bf 58}, 741 (1986).

\noindent
 {Thiffeault, J.-L.,  Chaos {\bf 20}, 017516 (2010).}
 
\noindent
Titov, V.S., Hornig, G., D{\'e}moulin, P.,  JGR {\bf 107}, 
A8 (2002).

\noindent
Tripathi, D., Mason H. E., Klimchuk, J.A.,  ApJ {\bf 723}, 713 (2010).

\noindent
Warren, H. P., Winebarger, A. R., Brooks, D.H.,  ApJ {\bf 711}, 228 (2010).

\noindent
Wilmot-Smith, A.L., Hornig, G., Pontin, D.I.,  ApJ {\bf 696}, 1339 (2009).

\noindent
Wilmot-Smith, A.L., Pontin, D.I., Hornig, G., 
 A\&A {\bf 516}, A5 (2010).

\noindent
 {Yeates, A.R., Hornig, G., J. Phys. A Math. Theor. {\bf 44}, 265501 (2011).}

\noindent
Yeates, A.R., Hornig, G., Wilmot-Smith, A.L.,  PRL {\bf 105} (8) 085002 (2010).

\end{document}